\documentclass[12pt]{iopart}
\usepackage{iopams}
\usepackage{color}
\usepackage{epsfig}   
\usepackage{graphics}
\newcommand{\be}{\begin{eqnarray}}
\newcommand{\ee}{\end{eqnarray}}

\begin{document}

\title[]{Early search for supersymmetric dark matter models at the LHC without missing energy}

\author{Joakim Edsj\"o, Erik Lundstr\"om\footnotemark[1], Sara Rydbeck\footnotemark[1], J\"orgen Sj\"olin}
\footnotetext[1]{Corresponding authors.}

\address{The Oskar Klein Centre for Cosmoparticle Physics, Department of Physics, Stockholm University, AlbaNova University Center, SE - 106 91 Stockholm, Sweden}

\begin{abstract}
We investigate early discovery signals for supersymmetry at the Large Hadron Collider without using information about missing transverse energy. Instead we use cuts on the number of jets and isolated leptons (electrons and/or muons). We work with minimal supersymmetric extensions of the standard model, and focus on phenomenological models that give a relic density of dark matter compatible with the WMAP measurements. An important model property for early discovery is the presence of light sleptons, and we find that for an integrated luminosity of only 200--300 pb$^{-1}$ at a center-of-mass energy of 10 TeV models with gluino masses up to $\sim 700$ GeV can be tested.
\end{abstract}

\eads{\mailto{edsjo@fysik.su.se}, \mailto{erik@fysik.su.se}, \mailto{sararyd@fysik.su.se}, \mailto{sjolin@fysik.su.se}}

\maketitle

%\tableofcontents

\section{Introduction} 
Many different cosmological observations unanimously confirm the existence of a neutral, non-baryonic matter component in the universe. The dark matter density needed is five times that of ordinary matter and according to the standard model of cosmology constitutes about 20\% of the energy content of the universe. However, current pieces of evidence stem from gravitational effects and the particle nature of the dark matter has yet to be established.

Particle astrophysics is an active and exciting field but methods of indirect and direct detection of dark matter particles present in our surroundings suffer from uncertainties due to astrophysics and the unknown distribution of the dark matter. 
In high energy collisions at particle accelerators we can hope to create the dark matter particles and in a more controlled environment determine its properties.

The CERN Large Hadron Collider (LHC) started up in November 2009. The plan is to let the proton beams collide at a center-of-mass energy of 7 TeV initially and if all goes well to ramp up the energy to 10 TeV, possibly via a series of intermediate steps with short periods of data taking. An integrated luminosity of at most a few hundred pb$^{-1}$ is expected to be reached during this first run. 

Among the various new physics phenomena that may show up one of the most well studied hypotheses is \emph{weak scale supersymmetry}. Not only does low energy supersymmetry offer a solution to the hierarchy problem but this type of models typically also provide a perfectly stable TeV-scale \emph{neutralino} -- a particle which only interacts via the weak force. In general any such Weakly Interacting Massive Particle (WIMP) is a good dark matter candidate since it automatically leaves a Big Bang relic density of the correct order to explain the currently measured value, and the neutralino is not an exception -- rather it is often used as a generic WIMP.

At the LHC, heavy supersymmetric particles may be produced, the strongly interacting gluinos and squarks being the most abundant, and subsequently decay into lighter states via cascade decay chains. At every step of the decay chain ordinary standard model particles will be produced and give rise to reconstructed jets and leptons in the data.\footnote{In this context, "leptons" refers to electrons and muons only.} In general, isolation of a signal coming from new physics is far from trivial due to the large hadronic standard model background present. Hard cuts need to be imposed on the final state to discriminate signal from background.

Many studies show that unlike the Higgs, superparticle states could be discovered early, i.e. at relatively low integrated luminosities. A stable WIMP would show up in the detector indirectly as \emph{missing transverse energy} (MET), and search strategies often make use of combinations of cuts on MET and a number of jets and/or leptons with high transverse momentum (see e.g. \cite{Baer:1995jetsMET,Baer:1995multilepton,Yamamoto:cuts}). In principle the MET background from standard model physics arises from the production of neutrinos and should be quite moderate.

Unfortunately, getting a handle on the MET measurement can be quite complicated \cite{ATLASexpperf}. Just to mention some of the dominating contributions, dead and noisy channels in the calorimeters, beam-gas interactions, and pure mis-measurements in cracks, inactive material and escaping muons can fake large missing energy in the detector. Due to the above reasons it might not be feasible to make use of the MET channel when analyzing the earliest data. On the other hand, more locally measured objects like muons, electrons and jets can be identified without requiring as much global understanding of the detector as MET. This makes them less difficult to calibrate and more suitable for early measurements.

Final states with many jets and leptons are rare in the standard model and as long as the jet and lepton multiplicities are large enough supersymmetry signal events should be visible above the background even without any requirements on MET, as shown in \cite{BaernoMET,Baer2008,Baer2009}. Such events are hence ideal to search for even before any extensive calibration of the detectors have taken place. 

In this paper we study the prospects of early LHC detection of supersymmetric models interesting from a dark matter perspective. We impose no cuts on the MET but instead require high lepton and jet multiplicities within the events. Compared to the work in \cite{BaernoMET,Baer2008,Baer2009}, where the constrained MSSM (or mSUGRA) was studied, we focus on the more phenomenological MSSM-7 supersymmetric framework, as well as an MSSM-8 extension. 
We also impose the constraint that the models can provide the correct relic density of neutralinos to explain all dark matter in the universe. Another difference with our study is that we use a different set of tools (which have different systematics).

To make predictions using only Monte Carlo simulations before any data has been taken is difficult for various reasons that will be discussed below. Our aim at this stage is to estimate which parts of parameter space could be probed during early data taking without cutting on missing energy but otherwise using cuts that are fairly standard. Our predictions are valid for a center-of-mass energy of 10 TeV but we also discuss how our results change if a first data sample at 7 TeV would turn out to be useful.

As knowledge of the detectors increases, other channels will become of high interest for supersymmetric dark matter studies. We wish to emphasize however that even after the detectors have been fully calibrated, channels without cuts on missing energy can provide valuable complementary information about the underlying physics model. This gives another reason for studying the prospects of detection of supersymmetry in channels not relying on missing energy. 

In Section 2 we give a short introduction to the dark matter problem. Section 3 defines the supersymmetric model that we are using for our phenomenological study.  The search strategy is described in Section 4 and the details of our analysis are given in Section 5. In Section 6 we present and discuss our results before concluding in Section 7.

%%%%%%%%%%
\section{Dark Matter}

The evidence for dark matter in the universe is overwhelming (see e.g.\ \cite{Bergstrom2000}) and suggests that the dark matter is made up by non-baryonic (elementary) particles. In particular, WMAP \cite{wmap5yr} has measured the dark matter density to be
\begin{equation}
\Omega_\chi h^2 = 0.1099 \pm 0.0062 ~(1\sigma),
\label{eq:wmap-oh2}
\end{equation}
where $\Omega_\chi$ is the dark matter density in units of the critical density and $h$ is the Hubble parameter in units of 100 km s$^{-1}$ Mpc$^{-1}$. In spite of the evidence, we still have not established the existence of dark matter other than through its gravitational interactions. There are however many hypothetical candidates for dark matter particles (see e.g.\ \cite{Bergstrom2009}) and one of the most promising class of hypotheses is Weakly Interacting Massive Particles (WIMPs), that naturally (due to the weak scale interaction cross section) have a relic density of roughly the right magnitude as measured by e.g.\ WMAP, Eq.~(\ref{eq:wmap-oh2}). A fascinating virtue of supersymmetric extensions of the standard model is that we get a WIMP dark matter candidate, the lightest neutralino, for free. Even if the neutralino naturally gives a relic density in the right ballpark, there is a large spread up and down depending on the details of the supersymmetric model. To be consistent with neutralinos making up all the dark matter of the Universe, we will therefore require that the lightest neutralino has a relic density within $2\sigma$ of the WMAP range, Eq.~(\ref{eq:wmap-oh2}). This effectively excludes part of the supersymmetric parameter space, where the annihilation cross section is either too large or too small.

Of course the neutralino does not need to make up all of the dark matter in the Universe, but it is much more appealing if we can explain the dark matter problem with just one new particle, especially one we get for free, as in supersymmetric models. One can also imagine that additional light states (that could arise in extensions of the MSSM models considered here) can affect the relic density of neutralinos via coannihilations. However, such light states will also affect the collider phenomenology studied here. To avoid these problems, we here stick to the simple scenario, where we assume that the neutralino makes up all of the dark matter, i.e.\ that the relic density satisfies Eq.~(\ref{eq:wmap-oh2}).

%%%%%%%%%%
\section{The Minimal Supersymmetric Standard Model} 

The R-parity conserving MSSM contains 105 new parameters and it is not feasible to allow for all of these to vary freely. Hence, one usually makes simplifying assumptions based on various schemes for supersymmetry breaking. A common assumption is that of cMSSM (or mSUGRA), where gauge coupling and mass unification is assumed at the grand unified (GUT) scale and the parameters are then run down to the electroweak scale. 

We here choose to work with more phenomenological models, the MSSM-7 and MSSM-8, where we instead give the parameters at the electroweak scale directly. Compared to cMSSM, this gives us a larger parameter space as we essentially have more freedom in the Higgs and neutralino/chargino sector. 
In the following subsections, we will briefly define our MSSM-7 and MSSM-8 models.

\subsection{Definition of the MSSM-7}

To reduce the 105 free parameters of the MSSM, a common assumption motivated by known low energy phenomenology is to set SUSY sources of CP violation to zero. Another is to assume the trilinear couplings and the soft sfermion mass matrices to be diagonal to avoid flavor changing neutral currents at tree level.

For our MSSM model, we will first use the phenomenological MSSM-7 model, as defined in Ref.~\cite{MSSM7}.

In this model, to reduce the number of parameters we make the approximation that only the third generation trilinear couplings $A_t$ and $A_b$ are non-zero and also introduce the common soft sfermion mass parameter $m_0$. In other words we take $\mathbf{A}_U={\rm diag}(0,0,A_t)$, $\mathbf{A}_D={\rm diag}(0,0,A_b)$, $\mathbf{A}_E=\mathbf{0}$ and $\mathbf{M}_Q=\mathbf{M}_U=\mathbf{M}_D=\mathbf{M}_L=\mathbf{M}_E=m_0\mathbf{1}$ following the notation of \cite{MSSM7}.

For simplicity, we also adopt the mSUGRA inspired 
GUT assumption for the gaugino mass parameters
\be  \label{eq:Mgut}
M_3 = {\alpha_s \over \alpha} \sin^2 \theta_{\rm W} M_2 = 
{3 \over 5} {\alpha_s \over \alpha} \cos^2 \theta_{\rm W} M_1,
\ee
where $\alpha$ and $\alpha_s$ are the fine-structure and strong coupling constants, respectively.
Approximately we have $2 M_1\sim M_2\sim 0.3 M_3$ and we choose to work with $M_2$ as our free parameter.

Through the minimization of the potential for the Higgs scalar fields the mass parameters and the bilinear coupling in the Higgs sector can be reduced into two independent parameters: the already mentioned $\tan \beta$ and the CP-odd Higgs boson mass $m_A$.

The seven free parameters left after these simplifying assumptions, $\mu$, $m_A$, $\tan \beta$, $M_2$, $m_0$, $A_b$ and $A_t$, defines the MSSM-7 reduction of the full MSSM parameter space \cite{MSSM7}. Note that all model parameters of the MSSM-7 are to be given at the electroweak scale. This is different to mSUGRA 
where most parameters are instead specified at the GUT scale and run down to low energies using the renormalization group equations.

\subsection{Definition of the MSSM-8}

As a consequence of the assumptions made in the MSSM-7 above it follows that all sfermions, except for the top squarks and possibly the bottom squarks and tau sleptons, typically become nearly degenerate in mass. This feature is in fact more generic in the MSSM-7 than it is in for example mSUGRA, where the running of the masses down from the GUT scale can introduce extra divergences.

While this mass degeneracy may be a convenient first assumption, it could also be overly restrictive when investigating properties that depend on the sfermion mass spectrum. A less degenerate spectrum opens up for new sparticle decay channels, and hence for new signal characteristics. Since we in this work especially care about the production of leptons within SUSY decay chains, it makes sense to try to relax the relation between the squark and slepton masses somewhat.

We therefore define the MSSM-8 as identical to the MSSM-7 except that we now split the common soft sfermion mass parameter $m_0$ into two: a common soft squark mass parameter $m_{\tilde{q}}$ and a common soft slepton mass parameter $m_{\tilde{l}}$. That is, we assume that $\mathbf{M}_Q=\mathbf{M}_U=\mathbf{M}_D=m_{\tilde{q}}\mathbf{1}$ and $\mathbf{M}_L=\mathbf{M}_E=m_{\tilde{l}}\mathbf{1}$. In total we are left with eight free parameters: $\mu$, $m_A$, $\tan \beta$, $M_2$, $m_{\tilde{q}}$, $m_{\tilde{l}}$, $A_b$ and $A_t$.

\section{SUSY search strategy}

In this work we set out to explore the prospects for detecting a signal from supersymmetry at the LHC, and the ATLAS detector \cite{ATLAS} in particular, through search channels not relying on MET. Instead we focus on the signal of multijets plus multileptons that can arise in the sparticle decay chains characteristic of many SUSY models but is harder to produce within the standard model.

With all the uncertainty that comes with a pre start-up simulation, the idea is to perform a quite conservative analysis. Specifically we keep our imposed sets of cuts simple and leave optimization until later, when the LHC has begun to collect some data reducing much of the uncertainty currently present within QCD phenomenology at these energies.

For consistency we choose to work with a single computational package, {\tt MadGraph/MadEvent} (MG/ME) \cite{MGME} interfaced with {\tt Pythia} \cite{PYTHIA} and the {\tt PGS} \cite{PGS} detector simulation, for all signal and background processes. We make efforts to tune the parameters in our analysis chain so that our results agree with those presented by the ATLAS collaboration \cite{ATLASnot10TeV}. We also investigate sources of systematic errors by varying the settings around our default values.

We analyze a couple of hundred models from wide scans over the SUSY parameter space, all required to pass numerous accelerator constraints and consistency requirements. An important restriction we impose on our models is that they have to provide the correct thermal relic abundance of neutralinos to account for all dark matter present in the universe. We calculate the neutralino relic density including coannihilations, threshold effects and resonances with {\tt DarkSUSY} \cite{DarkSUSY}. Note that this approach is different from that of \cite{BaernoMET,Baer2008,Baer2009} where no constraint on the relic density is imposed but focus instead is put on specific regions of the mSUGRA parameter space.

\subsection{Search channels}
\label{sec:searchchannels}

We choose a standard cut of four high energetic jets together with two isolated leptons, and furthermore divide the leptons into a same sign (SS) and an opposite sign (OS) search channel. In general both electrons and muons contribute to a dilepton signal but in case the identification of electrons within the early data turns out to be unreliable, as argued in \cite{Baer2008}, we also perform an analysis in which only muons are counted. 

A standard procedure to gain control over the background is to require the reconstructed leptons to carry a transverse momentum, $p_T$, exceeding 20 GeV. It can however be argued that this requirement could be overly restrictive when it comes to muons, for which a cut on $p_T>10$ GeV may be good enough \cite{Baer2008}. Which value (or combination of different values) is optimal of course remains to be seen once the collection of real data has started, and for comparison we here simply choose to analyze both cases for the dimuon channel. 

Besides the dilepton search we also consider the channel of four jets plus three leptons. While \cite{Baer2008,Baer2009} found this channel to be slightly less interesting for the very early data, it can still be used as a cross-check. 

In order to investigate what impact the required jet multiplicity has, we also consider three-jet channels corresponding to the searches mentioned above.

\subsection{Signal properties}

R-parity conservation ensures that every sparticle decay always results in the creation of an odd number of new lighter sparticles. Two-body sparticle decays always produce one sparticle plus one standard model particle while three-body decays may give rise to one sparticle plus two ordinary particles. The new sparticle can then further decay into yet another lighter sparticle plus one or several ordinary particles, and so on. The decay chain finally ends with the perfectly stable LSP (which constitutes a dark matter candidate).  The more steps in the chain, the more quarks and leptons can be produced and the clearer a signal we may get.

At hadron colliders the dominant SUSY production channels typically are into color charged sparticles, i.e. into gluinos, $\tilde{g}$, and/or squarks, $\tilde{q}$. Consequently, the most important production processes at the LHC should be: $pp\rightarrow \tilde{g}\tilde{g}$, $pp\rightarrow \tilde{g}\tilde{q}$ and $pp\rightarrow \tilde{q}\tilde{q}$ (where $\tilde{q}$ stands for any squark or anti-squark).

The gluinos and squarks can then decay into lighter neutralinos, $\tilde{\chi}_i^0$ ($i=1,2,3,4$), or charginos $\tilde{\chi}_i^\pm$ $(i=1,2)$ through processes like $\tilde{g}\rightarrow q\bar{q}\tilde{\chi}_i^0$, $\tilde{g}\rightarrow q\bar{q}'\tilde{\chi}_i^\pm$, $\tilde{q}\rightarrow q\tilde{\chi}_i^0$ and $\tilde{q}\rightarrow q'\tilde{\chi}_i^\pm$. The neutralinos and charginos could further decay into lighter neutralinos, charginos or sleptons according to e.g. $\tilde{\chi}_i^0\rightarrow Z^{(*)}\tilde{\chi}_j^0$, $\tilde{\chi}_i^0\rightarrow h\tilde{\chi}_j^0$, $\tilde{\chi}_i^0\rightarrow W^{\mp(*)}\tilde{\chi}_j^\pm$, $\tilde{\chi}_i^0\rightarrow \bar{l}\tilde{l}$, $\tilde{\chi}_i^\pm\rightarrow W^{\pm(*)}\tilde{\chi}_j^0$, $ \tilde{\chi}_i^\pm\rightarrow h\tilde{\chi}_j^\pm$, $\tilde{\chi}_i^\pm\rightarrow Z^{(*)}\tilde{\chi}_j^\pm$ or $\tilde{\chi}_i^\pm\rightarrow \bar{l}^{'}\tilde{l}$ (where $\tilde{l}$ denotes any slepton or anti-slepton), with possible slepton decay back into neutralinos or charginos in accordance with $\tilde{l}\rightarrow l\tilde{\chi}_j^0$ and $\tilde{l}\rightarrow l^{'}\tilde{\chi}_j^\pm$. Note that every one of the above mentioned decay processes can act as a source for jets and leptons reconstructed from the LHC data, either through quarks and charged leptons from direct emission or via the decay of the produced Higgs bosons, $h$, or gauge bosons, $Z$ and $W^\pm$.

The main factors determining the strength of a multijet plus multilepton signal are hence the gluino/squark production cross section as well as the structure of the spectrum of sparticles lighter than the gluinos/squarks. The production rate depends on the masses of the produced sparticles - the lighter the gluinos and/or the squarks are, the higher the production cross section generally is. However, a sparticle spectrum with favourable masses and mass splittings could also help increase the signal. For example, the presence of light sleptons in the model should enhance the number of leptons emitted during the decay chain.

With the above perspectives in mind we focus on models with light gluinos/squarks. To investigate the influence of the sparticle spectrum structure we analyze both MSSM-7 and MSSM-8 models with various sparticle mass spectra.

\subsection{Background properties}
\label{sec:backgroundproperties}
The standard model background can be severely constrained by demanding a large number of jets and leptons in the final state. The main background within the four jets plus two leptons channel comes from $t\bar{t}$ and $Z$ production \cite{Baer2009}, for each of which we produce large inclusive samples with up to two explicit hard jets. Here, and throughout the rest of this paper, $Z$ is used as shorthand notation for the full matrix element $\gamma^*/Z^{(*)}\rightarrow l^+l^-$.

In the four jets plus three leptons channel the $t\bar{t}Z$ background may also be of importance \cite{Baer2009} and we extensively simulate this process too. It should however be noted that due to practical restrictions on the maximum number of explicit jets feasible to include in the hard process within our analysis chain, our background simulation of the trilepton channel may not be as trustworthy as in the dilepton cases. (See Section \ref{sec:4j3l}.) This comes about since the dominating trilepton background seems to come from $Z$-events, for which an adequate number of explicit hard jets are hard to simulate. (See Section \ref{subsec:4j+2l}).

An important uncertainty in the determination of the background stems from the difficulty to correctly estimate the lepton \emph{fake rate}, i.e. the amount of leptons not originating from electroweak processes like $W^{\pm}$ and $Z$ decays. Fakes are typically produced in hadronization processes and do often propagate close to its parenting jet. This means that the fake rate is extremely sensitive to the imposed lepton isolation criteria (see Section \ref{sec:cuts}), and due to uncertainties present in the modeling of QCD physics it cannot be determined to any greater precision by other means than from the upcoming data itself. In addition there are also \emph{true fakes} arising from pure misidentification of non-leptons in the detector.

Opposite sign lepton pairs are extensively produced in electroweak processes during $t\bar{t}$ and $Z$ decays, and the OS dilepton search channel is not very much affected by the fake rate uncertainty. On the other hand same sign lepton pairs originating from $t\bar{t}$ and $Z$ production must contain at least one fake. This implies that our simulated SS dilepton background is rather uncertain, and before the results in this channel can be trusted the lepton fake rate really has to be measured in the data. Trilepton events from $t\bar{t}$ and $Z$ are of course also similarly affected by this fake rate uncertainty.

We note however that our analysis chain seems capable of producing lepton fake rates of the same order as those estimated by the ATLAS collaboration in \cite{ATLASnot10TeV}, something which adds some credibility to our SS dilepton background. For the pure SS dimuon channel, though, our fake rate seems to be roughly an order of magnitude below that of \cite{ATLASnot10TeV}.

The main contributions to the background for the three-jet channels should also come from $t\bar{t}$, $Z$ and $t\bar{t}Z$.

\section{Analysis}
\label{sec:analysis}
We analyze the detection prospects for 100 MSSM-7 and 100 MSSM-8 models scattered over the ($m_{squark}$,$m_{gluino}$)-plane.\footnote{Here $m_{squark}$ sets out to represent a generic squark mass scale, for which we take the up-squark mass $m_{\tilde{u}_L}$ (which is also typically close to the model input parameter $m_{\tilde{q}}$) as a good representative. Subsequent plots of the $(m_{squark},m_{gluino})$-plane hence actually show the physical $(m_{\tilde{u}_L},m_{\tilde{g}})$-plane.} The models are extracted from extensive scans over the MSSM parameter space and all satisfy $m_{\tilde{q}}<1200$ GeV and $m_{\tilde{g}}<1000$ GeV, pass a large variety of accelerator constraints and provide a relic density of neutralino dark matter within 2$\sigma$ of the WMAP range, Eq.~(\ref{eq:wmap-oh2}). For the MSSM-7, we have selected a set of typical representative models in various mass ranges, whereas for MSSM-8, we have focused on models with light sleptons.

Within the MSSM-7, models with many different properties are chosen in order to search for correlations with the signal. Partly motivated by the discussion in \cite{Baer2008}, the mass splittings between the light squarks and the stops, sbottoms and staus, as well as between the neutralinos, are varied. Various values of $\tan\beta$ and $\mu$ are investigated. We also examine models where all neutralinos and charginos are lighter than the produced gluinos and squarks, as this gives many decay opportunities and hence possibly a larger signal.

The chosen MSSM-8 models all include light sleptons, typically within the 100-200 GeV mass range. In addition, we require all neutralinos and charginos to be lighter than the gluinos and squarks. These requirements open up for decays via the sleptons and hence a larger rate of production of final state leptons.

For the signal we consider production of all possible two-body combinations of gluinos and (anti-)squarks. In total this adds up to 325 different hard process final states to simulate. We assume the production cross section into other states including neutralinos, charginos and sleptons to be sufficiently low to be safely neglected for our purposes. For each model we simulate $250\ \!000$ signal events.

For the background, matched samples of $t\bar{t}+0,1,2$ jets, $Z+0,1,2$ jets and $t\bar{t}Z$ are simulated and, in the cases of the $t\bar{t}$ and $Z$ samples, normalized in agreement with NLO calculations \cite{MCFM}. In total we generate $12\cdot10^6$ $t\bar{t}$ events ($\sim 30$ fb$^{-1}$), $26\cdot10^6$ $Z$ events ($\sim 10$ fb$^{-1}$), and $6\cdot10^6$ $t\bar{t}Z$ events ($\sim10^5$ fb$^{-1}$).

The signal and background samples are passed through a simplified model of the ATLAS detector in which jets and leptons are reconstructed, whereafter we impose cuts appropriate for our analysis. Main focus is put on the four-jet plus dilepton search channels but also trilepton as well as three-jet channels are examined.

For each model we estimate the minimum integrated luminosity required for detection at the LHC by demanding
\be
\label{eq:detectionlimit}
S \geq \max \{5\sqrt{B},5\},
\ee
where $S$ and $B$ are the post-cuts number of expected signal and background events, respectively. That is, we require a $5 \sigma$ significance but also at least five signal events.

%%%%%%%%%%
\subsection{Tools}
The MSSM particle spectrum and relic density calculations are carried out with {\tt DarkSUSY-5.0.4} \cite{DarkSUSY}, a comprehensive numerical package for neutralino dark matter calculations. {\tt DarkSUSY} also imposes a large variety of accelerator constraints on the models, including bounds on sparticle and higgs boson masses.
The decay widths of the sparticles are calculated with {\tt SDECAY-1.1a} \cite{sdecay}.

For the event generation of the standard model background as well as the supersymmetry signal we use {\tt MadGraph/MadEvent-4.4.21} ({\tt MG/ME}) \cite{MGME}. The {\tt MG/ME} package includes a number of interfaced programs: {\tt MadGraphII} identifies the relevant tree-level diagrams for the hard process and finds analytical expressions for the production cross section, while {\tt MadEvent} constitutes the parton level Monte Carlo event generator. The events are then fed into {\tt Pythia-PGS-2.0.27} via {\tt MG/ME}'s interface that includes the possibility of jet matching for generation of inclusive samples. Except for $\tau$-decays, which is handled by {\tt Tauola}, {\tt Pythia} takes care of the parton showering, cascade decays and hadronization. In the last step, the events are passed through {\tt PGS}, the Pretty Good Simulation code which sets out to be a fast, but still fairly accurate, detector simulation.

\subsection{Settings}
We consider proton-proton collisions at a 10 TeV center-of-mass energy and choose {\tt cteq6l1} \cite{cteq6} for the parton distribution function.

We make use of {\tt MG/ME}'s option of variable renormalization and factorization scales, and for our standard model background processes with explicit extra jets in the final states our choice of matching procedure is the $K_T$-jet MLM scheme \cite{jetmatching}.

In {\tt Pythia} string fragmentation according to the Lund model is used. We include initial and final state radiation but no multiple interactions. The jet measure cutoff used in the matching scheme is set to QCUT=25 GeV.

For {\tt PGS} we use parameters tuned to mimic the ATLAS detector, as provided in {\tt MadGraph/MadEvent-4.4.21}. There are 81$\times$63 calorimeter cells spanning the $\eta\times\phi$ (pseudorapidity $\times$ azimuthal angle) plane with $\Delta\eta=0.1$ and full coverage in $\phi$. The electromagnetic and hadronic calorimeter resolutions are taken to be $10\%/\sqrt{E}$ and $80\%/\sqrt{E}$, respectively. A magnetic field of 2 T is assumed and tracking is covered for $|\eta|<2.5$. Electron and muon coverage are $|\eta|<3.0$ and $|\eta|<2.4$, respectively.
Jets are reconstructed using a cone algorithm with cluster finder cone size $\Delta R=0.4$.\footnote{$\Delta R=\sqrt{(\eta_1-\eta_2)^2+(\phi_1-\phi_2)^2}$}

By comparing the lepton efficiencies for a $W^{\pm}$-sample using our analysis chain with those presented by the ATLAS collaboration \cite{ATLASnot10TeV} we find that an imposed additional efficiency, on top of what comes out from our detector simulation, can lead to an even better agreement. For this reason, we explicitly apply a muon efficiency of 0.9 for all $|\eta|$ and an electron efficiency of 0.85 when $|\eta|<1$ and 0.65 otherwise.

\subsection{Cuts}
\label{sec:cuts}
After an event has passed through the detector simulation we apply our own cuts, both in order to better trust the result of the detector response and to gain in signal over background.

We only consider isolated leptons, meaning that they could not easily have originated in jets.
In PGS, electrons are isolated by default but not muons. The requirements on electrons are that the transverse calorimeter energy in an ($3\times3$) cell grid around the electron, excluding the cell with the electron, in total has to be less than 10\% of the electron's $E_T$ and that the summed $p_T$ of tracks within a $\Delta R=0.4$ cone around the electron, again excluding the electron itself, must be less than 5 GeV. Also, 
the ratio of the calorimeter cell energy to the $p_T$ has to be between 0.5 and 1.5. For muons  
we require the summed $p_T$, excluding the muon itself, within a $\Delta R=0.4$ cone around the muon
to be less than 10 GeV. 

In addition we require any lepton (electron or muon) to be separated from the nearest jet by at least  $\Delta R=0.3$ and to have a minimum $p_T$ of 20 GeV. In the search channel where we consider only muons and not electrons, we also let the minimum $p_T$ be 10 GeV following \cite{Baer2008,Baer2009}. While lowering the $p_T$ cut could potentially cause more muons from leptonically decaying b-jets to be isolated, there is still much room for tighter isolation criteria at low $p_T$, using both tracks and calorimeter cells, without significant loss of efficiency. Since an accurate prediction using such criteria is difficult to make without a control sample from real data, we choose to leave them out in the present analysis.

We ignore electrons within $1.37\le|\eta|\le1.52$ since the detector response is unreliable in that region \cite{ATLASnot10TeV}.

In order to reduce the $Z$-background we cut away events where any pair of same flavour, opposite sign leptons has an invariant mass in the range $75$ GeV$\le m(l^+l^-)\le 105$ GeV.

In the four-jet channels we require the leading jet to carry a transverse momentum $p_T>100$ GeV while each of the three sub-leading jets must satisfy $p_T>50$ GeV, following the procedure of many supersymmetry searches. In the three-jet channels we drop the leading jet and only require three jets with $p_T>50$ GeV each.

For the channels with only muons instead of leptons, for the eventuality that electron identification cannot be trusted, we count the electrons 
reconstructed by the detector simulation as jets. 

Since we are considering the prospects of early discovery we do not apply a transverse sphericity cut, the effect of which would drown in jet systematics at this point. (See Section \ref{sec:systematicuncertainties}.)

We do not apply any cut on missing transverse energy.

\section{Results}
%==============================================================================
\begin{figure}
        \centering
\includegraphics[scale=0.48]{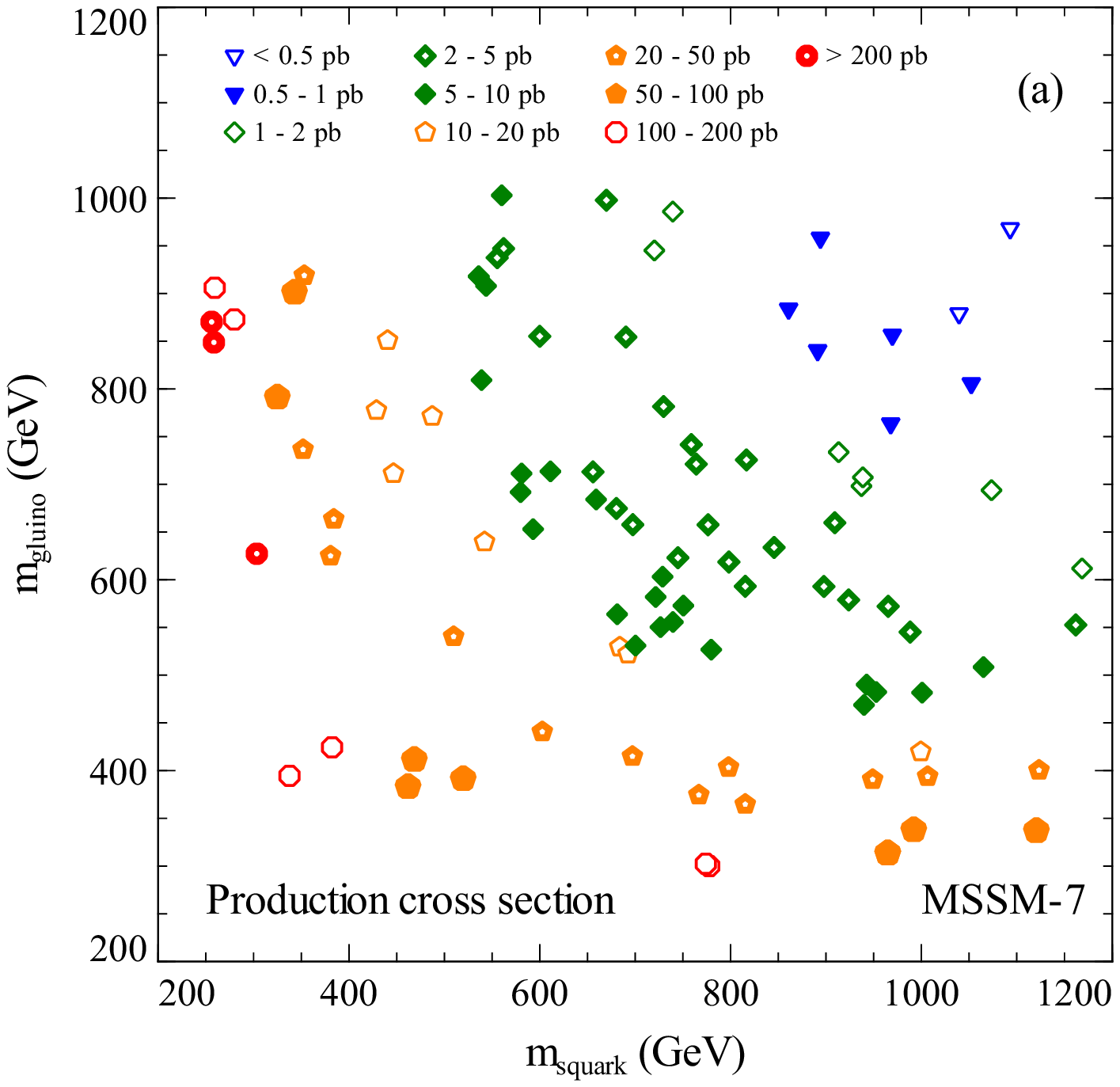} \hspace{0.5cm}
\includegraphics[scale=0.48]{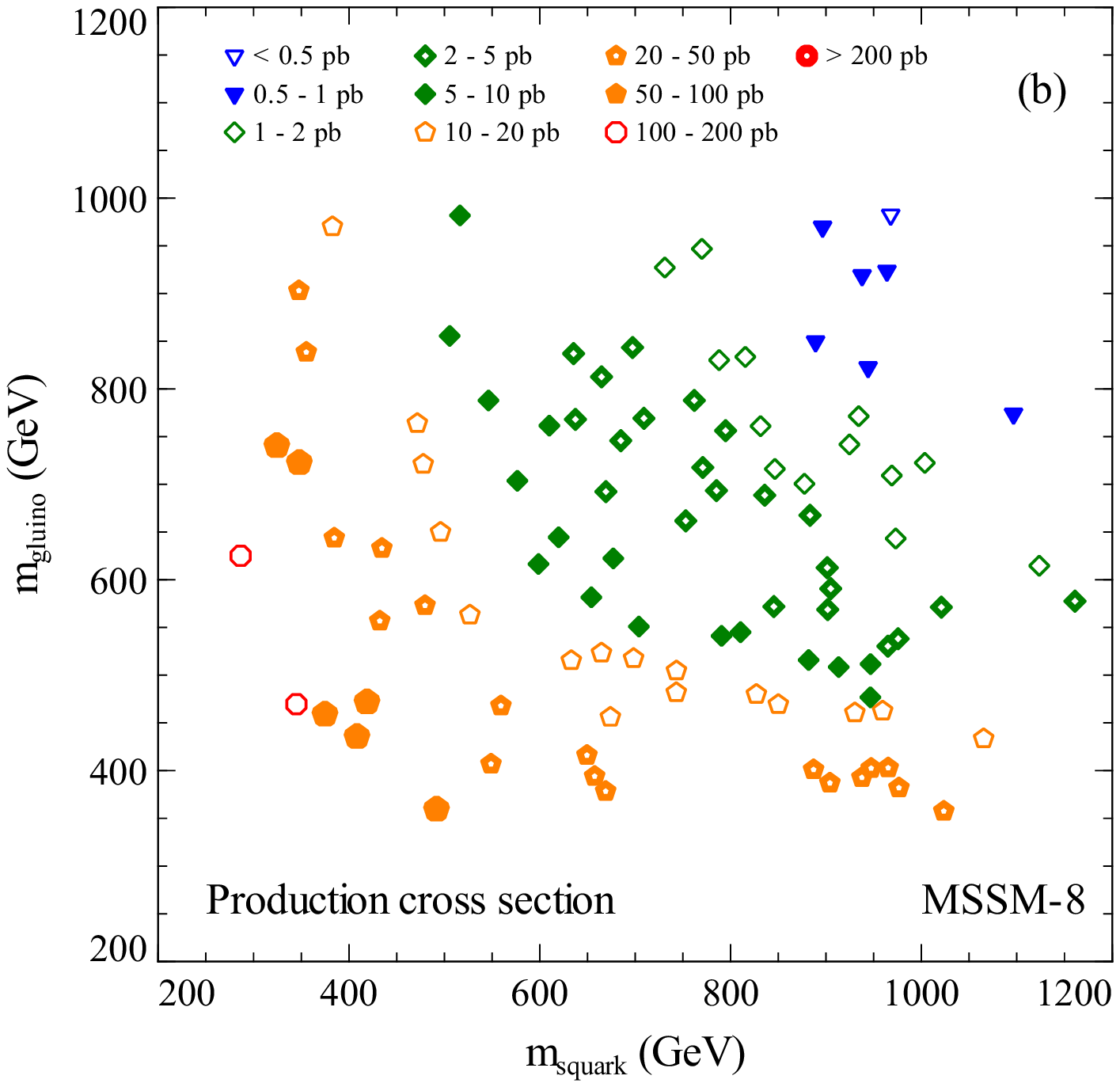}
 \caption{Total gluino/squark production cross sections for representative sets of (a) MSSM-7 and (b) MSSM-8 models. The various symbols correspond to different ranges for the value of the cross section.}
        \label{fig:xsec}
\end{figure}
%==============================================================================
To get the reader oriented with the models under study we plot their positions in the $(m_{squark},m_{gluino})$-plane in Figure \ref{fig:xsec}. We divide the points into two plots, one for the MSSM-7 set of models and one for the MSSM-8 set. For each model the summed cross section for pair production of any combination of gluinos and/or (anti-)squarks is indicated. As expected, the gluino/squark production cross section shows up as strongly anti-correlated with the gluino/squark masses. The results for the MSSM-7 and the MSSM-8 are very similar - the slepton masses do not affect the production of gluinos and squarks.

In addition to our wider MSSM scans we choose three pairs of benchmark models for a closer investigation of their properties. Each pair consists of one MSSM-7 and one MSSM-8 model that differ only by the value of the slepton mass parameter $m_{\tilde{l}}$. The various pairs are otherwise spread out in the $(m_{squark},m_{gluino})$-plane and in the MSSM parameter space in general. The benchmark model parameters are listed in Table \ref{table:benchmarkmodels} together with some selected sparticle masses, the relic density and the total gluino/squark production cross section. Note that model {\bf B7} gives a too high relic abundance of neutralinos, and is hence excluded on cosmological grounds. This one model is included in our analysis despite this fact, in the purpose of better understanding the effect that light sleptons 
have for the LHC signal.

%==============================================================================
\begin{table}
\begin{center}
\begin{tabular}{| c || c | c || c | c || c | c |}
\hline
{\bf Model} & {\bf A7} & {\bf A8} & {\bf B7} & {\bf B8} & {\bf C7} & {\bf C8} \\
\hline
\hline
$\mu$ & \multicolumn{2}{c||}{$130$} & \multicolumn{2}{c||}{$290$} & \multicolumn{2}{c|}{$360$} \\
$m_A$ & \multicolumn{2}{c||}{$700$} & \multicolumn{2}{c||}{$1000$} & \multicolumn{2}{c|}{$600$} \\
$\tan\beta$ & \multicolumn{2}{c||}{$15$} & \multicolumn{2}{c||}{$5$} & \multicolumn{2}{c|}{$50$} \\
$M_2$ & \multicolumn{2}{c||}{$190$} & \multicolumn{2}{c||}{$170$} & \multicolumn{2}{c|}{$115$} \\
$m_{\tilde{q}}$ & \multicolumn{2}{c||}{$380$} & \multicolumn{2}{c||}{$650$} & \multicolumn{2}{c|}{$660$} \\
\hline
$m_{\tilde{l}}$ & $380$ & $180$ & $650$ & $110$ & $660$ & $260$ \\
\hline
$A_b/m_{\tilde{q}}$ & \multicolumn{2}{c||}{$2.5$} & \multicolumn{2}{c||}{$-1.5$} & \multicolumn{2}{c|}{$0$} \\
$A_t/m_{\tilde{q}}$ & \multicolumn{2}{c||}{$0$} & \multicolumn{2}{c||}{$2.5$} & \multicolumn{2}{c|}{$0.9$} \\
\hline
\hline
$m_{\tilde{g}}$ & \multicolumn{2}{c||}{$658$} & \multicolumn{2}{c||}{$589$} & \multicolumn{2}{c|}{$399$} \\
$m_{\tilde{u}_L}$ & \multicolumn{2}{c||}{$376$} & \multicolumn{2}{c||}{$648$} & \multicolumn{2}{c|}{$658$} \\
$m_{\tilde{b}_1}$ & \multicolumn{2}{c||}{$376$} & \multicolumn{2}{c||}{$642$} & \multicolumn{2}{c|}{$591$} \\
$m_{\tilde{t}_1}$ & \multicolumn{2}{c||}{$413$} & \multicolumn{2}{c||}{$424$} & \multicolumn{2}{c|}{$602$} \\
\hline
$m_{\tilde{e}_L}$ & $382$ & $185$ & $651$ & $118$ & $661$ & $264$ \\
$m_{\tilde{\mu}_L}$ & $382$ & $185$ & $651$ & $117$ & $660$ & $260$ \\
$m_{\tilde{\tau}_1}$ & $378$ & $176$ & $649$ & $107$ & $637$ & $194$ \\
\hline
$m_{\chi^0_1}$ & \multicolumn{2}{c||}{$71$} & \multicolumn{2}{c||}{$80$} & \multicolumn{2}{c|}{$57$} \\
$m_{\chi^0_2}$ & \multicolumn{2}{c||}{$120$} & \multicolumn{2}{c||}{$147$} & \multicolumn{2}{c|}{$109$} \\
$m_{\chi^0_3}$ & \multicolumn{2}{c||}{$143$} & \multicolumn{2}{c||}{$294$} & \multicolumn{2}{c|}{$368$} \\
$m_{\chi^0_4}$ & \multicolumn{2}{c||}{$236$} & \multicolumn{2}{c||}{$327$} & \multicolumn{2}{c|}{$377$} \\
$m_{\chi^\pm_1}$ & \multicolumn{2}{c||}{$101$} & \multicolumn{2}{c||}{$144$} & \multicolumn{2}{c|}{$108$} \\
$m_{\chi^\pm_2}$ & \multicolumn{2}{c||}{$236$} & \multicolumn{2}{c||}{$324$} & \multicolumn{2}{c|}{$379$} \\
\hline
\hline
$\Omega_\chi h^2$ & $0.12$ & $0.10$ & $3.8$ & $0.11$ & $0.11$ & $0.10$ \\
\hline
\hline
$\sigma_{prod}$ (pb) & \multicolumn{2}{c||}{$35.0$} & \multicolumn{2}{c||}{$8.6$} & \multicolumn{2}{c|}{$34.7$} \\
\hline
\end{tabular}
\end{center}
\caption{\label{table:benchmarkmodels} Benchmark models with derived sparticle masses, dark matter relic density and total gluino/squark production cross section. Each pair contains one MSSM-7 and one MSSM-8  model for which only the slepton mass parameter $m_{\tilde{l}}$ differs. All quantities with mass dimension are given in units of GeV.}
\end{table}
%==============================================================================

By performing cuts on the simulated data, the background and signal cross sections for each search channel are determined. The standard model background cross sections are summarized in Table \ref{bgtable}, together with the statistical Poisson uncertainty from our simulations. In the same way, the signal cross sections for our benchmark models are presented in Table \ref{sigtable}.

%==============================================================================
\begin{table}
\begin{center}
\begin{tabular}{|l|c|c|c|c|c|c|c|c|}
\hline
\bf{Process}&\multicolumn{1}{|c|}{$t\bar{t}$ (fb)}&\multicolumn{1}{|c|}{$Z$ (fb)}&\multicolumn{1}{|c|}{$t\bar{t}Z$ ($10^{-3}$fb)}&\multicolumn{1}{|c|}{{\bf Total} (fb)}\\
\hline
\hline
Channel&\multicolumn{4}{|c|}{$4j+$leptons of $p_T>20$GeV}\\
\hline
\hspace{0.4cm}$SS$ &$11.2\pm 0.6$ &0.3$\pm0.2$ &$144.3\pm0.9 $&$11.6\pm0.6$ \\
\hspace{0.4cm}$OS$&$132\pm 2$&$7.4\pm0.9 $ &$281\pm1 $ &$140\pm2 $ \\
\hspace{0.5cm}$3l$ &$0.16\pm 0.07$& $0.5\pm0.2 $& $26.1\pm0.4 $& $0.7\pm0.3 $ \\
\hline
\hline
Channel&\multicolumn{4}{|c|}{$4j+$muons of $p_T>20$GeV}\\
\hline
\hspace{0.4cm}$SS$ &$0.22\pm0.08 $ &$0$ $(<0.1)$&$37.3\pm0.5$&$0.26\pm0.08$ \\
\hspace{0.4cm}$OS$&$36\pm1$&$4.6\pm0.7$&$128.6\pm0.9 $&$40\pm1$ \\
\hspace{0.5cm}$3\mu$ &$0$  $(<0.006)$& $0$ $(<0.1)$& $5.0\pm0.2$& $(5.0\pm0.2)\cdot10^{-3} $ \\
\hline
\hline
Channel&\multicolumn{4}{|c|}{$4j+$muons of $p_T>10$GeV}\\
\hline
\hspace{0.4cm}$SS$&$3.1\pm0.3 $&$0$ $(<0.1)$&$49.6\pm0.5$&$3.2\pm0.3$ \\
\hspace{0.4cm}$OS$&$51\pm1$ &$7.4\pm0.9$&$199\pm1 $&$59\pm2 $ \\
\hspace{0.5cm}$3\mu$ &$0.09\pm0.05$& $0$ $(<0.1)$& $9.6\pm0.2$& $0.10\pm0.2 $ \\
\hline
\hline
Channel&\multicolumn{4}{|c|}{$3j+$leptons of $p_T>20$GeV}\\
\hline
\hspace{0.4cm}$SS$ & $45\pm 1$ &$1.6\pm0.4 $ &$475\pm2 $&$47\pm1$ \\
\hspace{0.4cm}$OS$ & $670\pm 5$ &72$\pm3 $ &$812\pm2 $ &$743\pm5 $ \\
\hspace{0.5cm}$3l$ &1.7$\pm0.2 $& $2.3\pm0.5 $ &$95.3\pm0.7 $ &$4.1\pm0.6$ \\
\hline
\hline
Channel&\multicolumn{4}{|c|}{$3j+$muons of $p_T>20$GeV}\\
\hline
\hspace{0.4cm}$SS$& $1.6\pm0.2$ &$0$ $(<0.1)$ &$129.3\pm0.9$&$1.7\pm0.2$ \\
\hspace{0.4cm}$OS$ & $173\pm2$ &$45\pm2$ &$351\pm1 $&$219\pm3 $ \\
\hspace{0.5cm}$3\mu$ &$0$  $(<0.006)$& $0$ $(<0.1)$ &$18.6\pm0.3$ &$(18.6\pm0.3)\cdot10^{-3} $ \\
\hline
\hline
Channel&\multicolumn{4}{|c|}{$3j+$muons of $p_T>10$GeV}\\
\hline
\hspace{0.4cm}$SS$& $14.9\pm0.7$ &$0$ $(<0.1)$ &$168\pm1$&$15.1\pm0.7$ \\
\hspace{0.4cm}$OS$ & $253\pm3$ &$73\pm3$ &$546\pm 2$&$326\pm4 $ \\
\hspace{0.5cm}$3\mu$ &$0.09\pm0.05$& $0.1\pm0.1$&$37.3\pm0.5 $&$0.2\pm0.1 $ \\
\hline

\end{tabular}
\end{center}
\caption{\label{bgtable} Background cross sections for different search channels. The indicated errors are statistical. 
In the cases where our simulations have not generated a single event that pass the cuts, we quote an upper limit (set by assuming one event, given the generated luminosity) within parentheses.}
\end{table}
%==============================================================================

%==============================================================================
\begin{table}
\begin{center}
\begin{tabular}{|l|c|c|c|c|c|c|c|c|c|c|}
\hline
\bf{Model}&\multicolumn{1}{|c|}{{\bf A7} (fb)}&\multicolumn{1}{|c|}{{\bf A8} (fb)}&\multicolumn{1}{|c|}{{\bf B7}  (fb)}&\multicolumn{1}{|c|}{{\bf B8}  (fb)}&\multicolumn{1}{|c|}{{\bf C7}  (fb)}&\multicolumn{1}{|c|}{{\bf C8}  (fb)}\\
\hline
\hline
Channel&\multicolumn{6}{|c|}{$4j+$leptons of $p_T>20$GeV}\\
\hline
\hspace{0.4cm}$SS$&$30\pm2$&$63\pm3$&$21\pm0.9$&$148\pm2$&$30\pm2$&$46\pm3$\\
\hspace{0.4cm}$OS$&$87\pm4$&$200\pm5$&$61\pm1$&$256\pm3$&$107\pm4$&$266\pm6$\\
\hspace{0.5cm}$3l$&$4.6\pm0.8$&$24\pm2$&$5.0\pm0.4$&$40\pm1$&$3.6\pm0.7$&$14\pm1$\\
\hline
\hline
Channel&\multicolumn{6}{|c|}{$4j+$muons of $p_T>20$GeV}\\
\hline
\hspace{0.4cm}$SS$ &$8\pm1$&$19\pm2$&$5.8\pm0.4$&$45\pm1$&$10\pm1$&$14\pm1$\\
\hspace{0.4cm}$OS$&$ 36\pm2$&$101\pm4$&$27\pm1$&$110\pm2$&$53\pm3$&$145\pm5$\\
\hspace{0.5cm}$3\mu$&$0.4\pm0.2$&$5.2\pm0.9$&$0.9\pm0.2$&$8.5\pm0.5$&$1.2\pm0.4$&$3.9\pm0.7$\\
\hline
\hline
Channel&\multicolumn{6}{|c|}{$4j+$muons of $p_T>10$GeV}\\
\hline
\hspace{0.4cm}$SS$ &$20\pm2$&$33\pm2$&$8.9\pm0.6$&$57\pm1$&$15\pm1$&$23\pm2$\\
\hspace{0.4cm}$OS$&$ 72\pm3$&$160\pm5$&$42\pm1$&$152\pm2.$&$108\pm4$&$302\pm7$\\
\hspace{0.5cm}$3\mu$&$4.2\pm0.8$&$14\pm1$&$1.5\pm0.2$&$15.5\pm0.7$&$4.0\pm0.7$&$11\pm1$\\
\hline
\hline
Channel&\multicolumn{6}{|c|}{$3j+$leptons of $p_T>20$GeV}\\
\hline
\hspace{0.4cm}$SS$ &$65\pm3$&$142\pm5$&$31\pm1$&$229\pm3$&$46\pm3$&$76\pm3$\\
\hspace{0.4cm}$OS$&$187\pm5$&$457\pm8$&$91\pm2$&$408\pm4$&$161\pm5$&$394\pm7$\\
\hspace{0.5cm}$3\mu$&$11\pm1$&$54\pm3$&$8.0\pm0.5$&$67\pm2$&$7\pm1$&$22\pm2$\\
\hline
\hline
Channel&\multicolumn{6}{|c|}{$3j+$muons of $p_T>20$GeV}\\
\hline
\hspace{0.4cm}$SS$ &$19\pm2$&$41\pm2$&$8.6\pm0.5$&$68\pm2$&$15\pm1$&$21\pm2$\\
\hspace{0.4cm}$OS$&$73\pm3$&$225\pm6$&$38\pm1$&$169\pm2$&$81\pm3$&$213\pm5$\\
\hspace{0.5cm}$3\mu$&$2.2\pm0.6$&$13\pm1$&$1.4\pm0.2$&$14.8\pm0.7$&$1.5\pm0.5$&$6.3\pm0.9$\\
\hline
\hline
Channel&\multicolumn{6}{|c|}{$3j+$muons of $p_T>10$GeV}\\
\hline
\hspace{0.4cm}$SS$ &$42\pm2$&$73\pm3$&$12.7\pm0.7$&$86\pm2$&$23\pm2$&$36\pm2$\\
\hspace{0.4cm}$OS$&$152\pm5$&$347\pm7$&$58\pm1$&$231\pm3$&$157\pm5$&$434\pm8$\\
\hspace{0.5cm}$3\mu$&$9\pm1$&$32\pm2$&$2.6\pm0.3$&$27\pm1$&$5.8\pm0.9$&$18\pm2$\\
\hline

\end{tabular}
\end{center}
\caption{\label{sigtable} Signal cross sections for different search channels for the benchmark models defined in Table \ref{table:benchmarkmodels}. The indicated errors are statistical.}
\end{table}
%==============================================================================

Using Eq.~(\ref{eq:detectionlimit}) we can then predict, for each 
MSSM model in our scan, the minimum amount of integrated luminosity needed for detection of a signal against the standard model background in the ATLAS detector for proton-proton collisions at a center-of-mass energy of 10 TeV. On doing so we neglect the uncertainty in the background but propagate the statistical errors in the signal cross section. The resulting statistical errors for the derived integrated luminosities are typically small ($\lesssim 20\%$). The reaches in various search channels are presented in Figures \ref{fig:4j2l}--\ref{fig:3j2l} and the results are discussed in more detail in Sections \ref{sec:4j2l}--\ref{sec:3jl}.

\begin{figure}
        \centering
\includegraphics[scale=0.48]{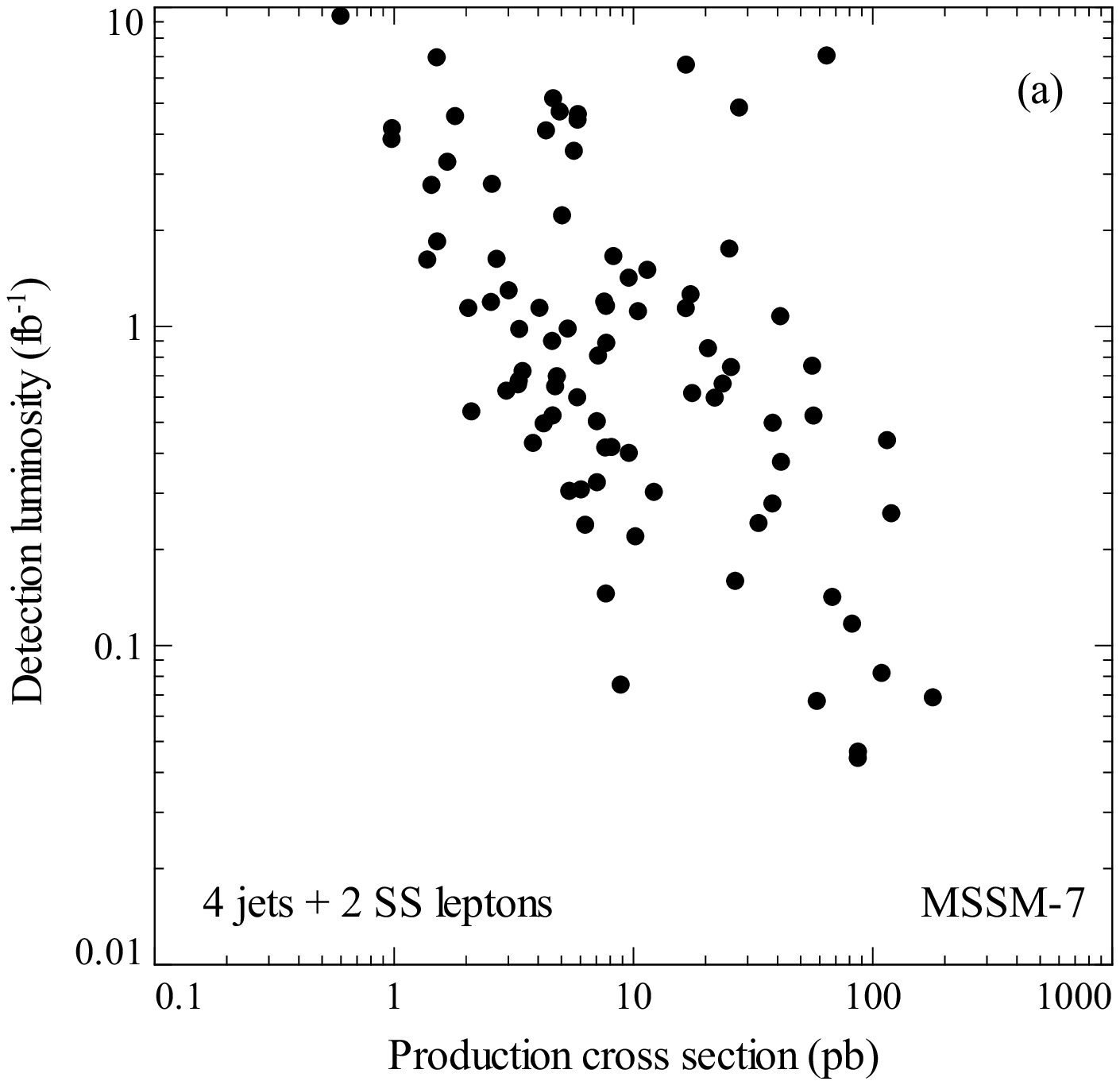}
\includegraphics[scale=0.48]{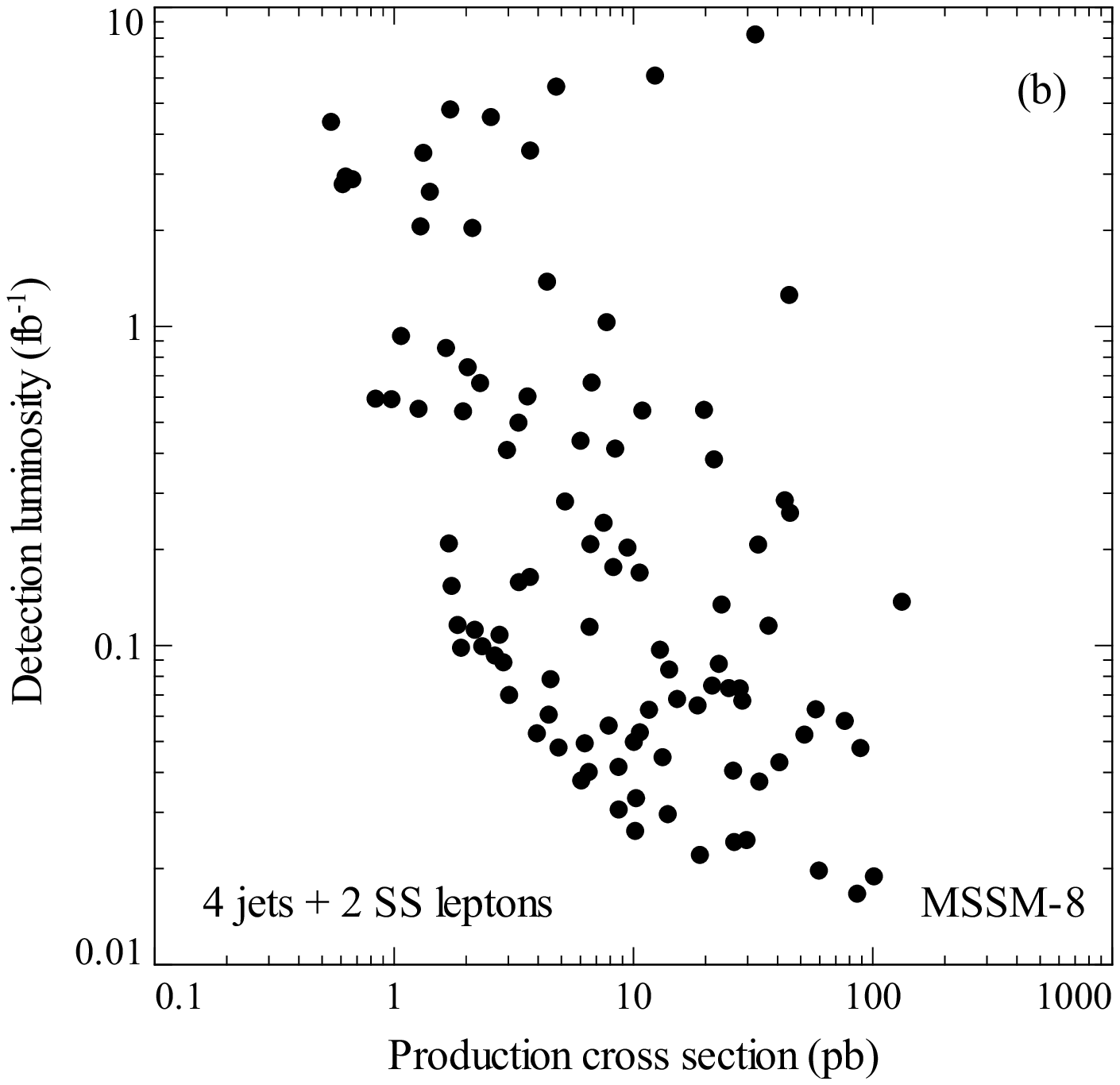}

\includegraphics[scale=0.48]{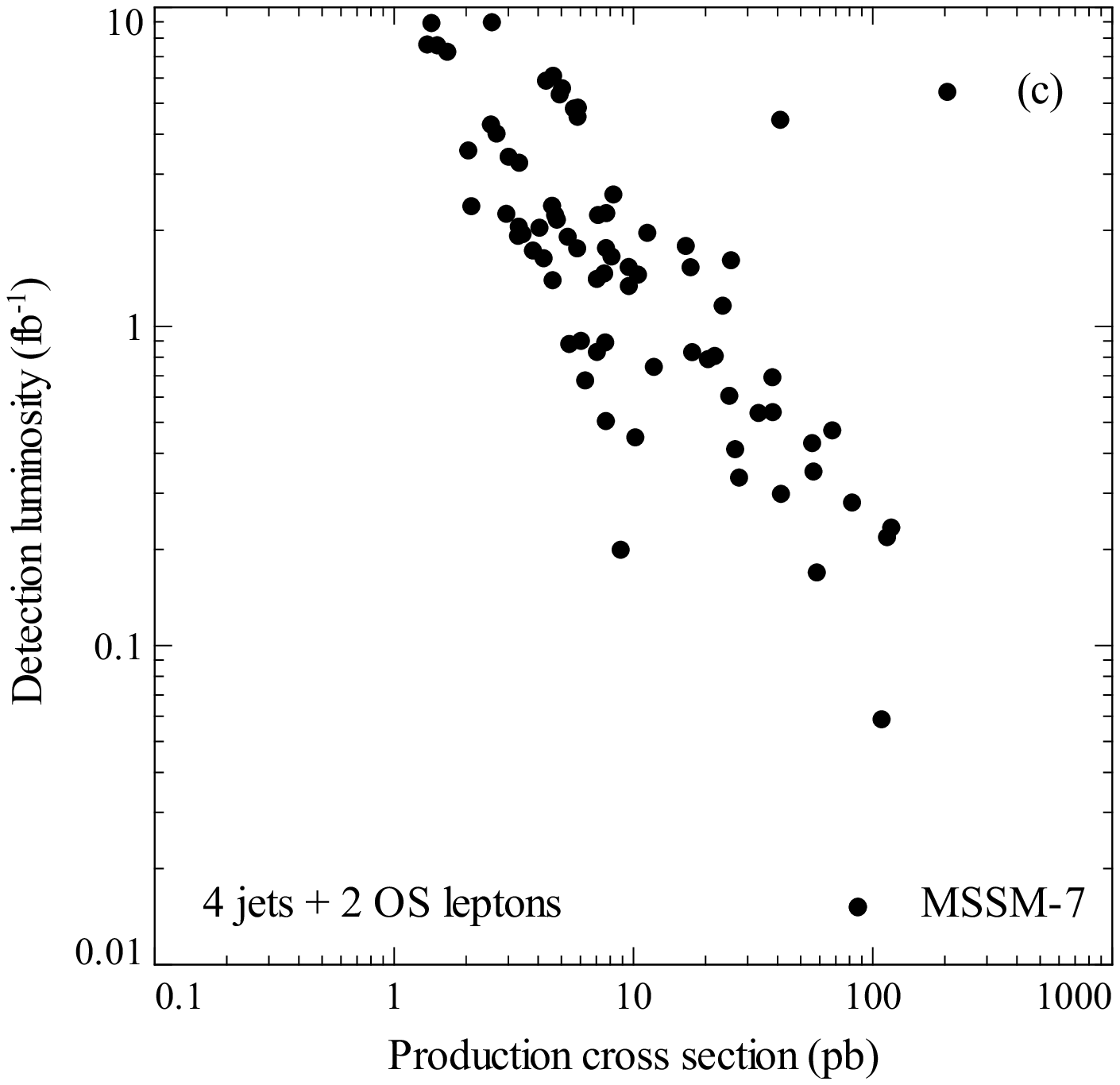}
\includegraphics[scale=0.48]{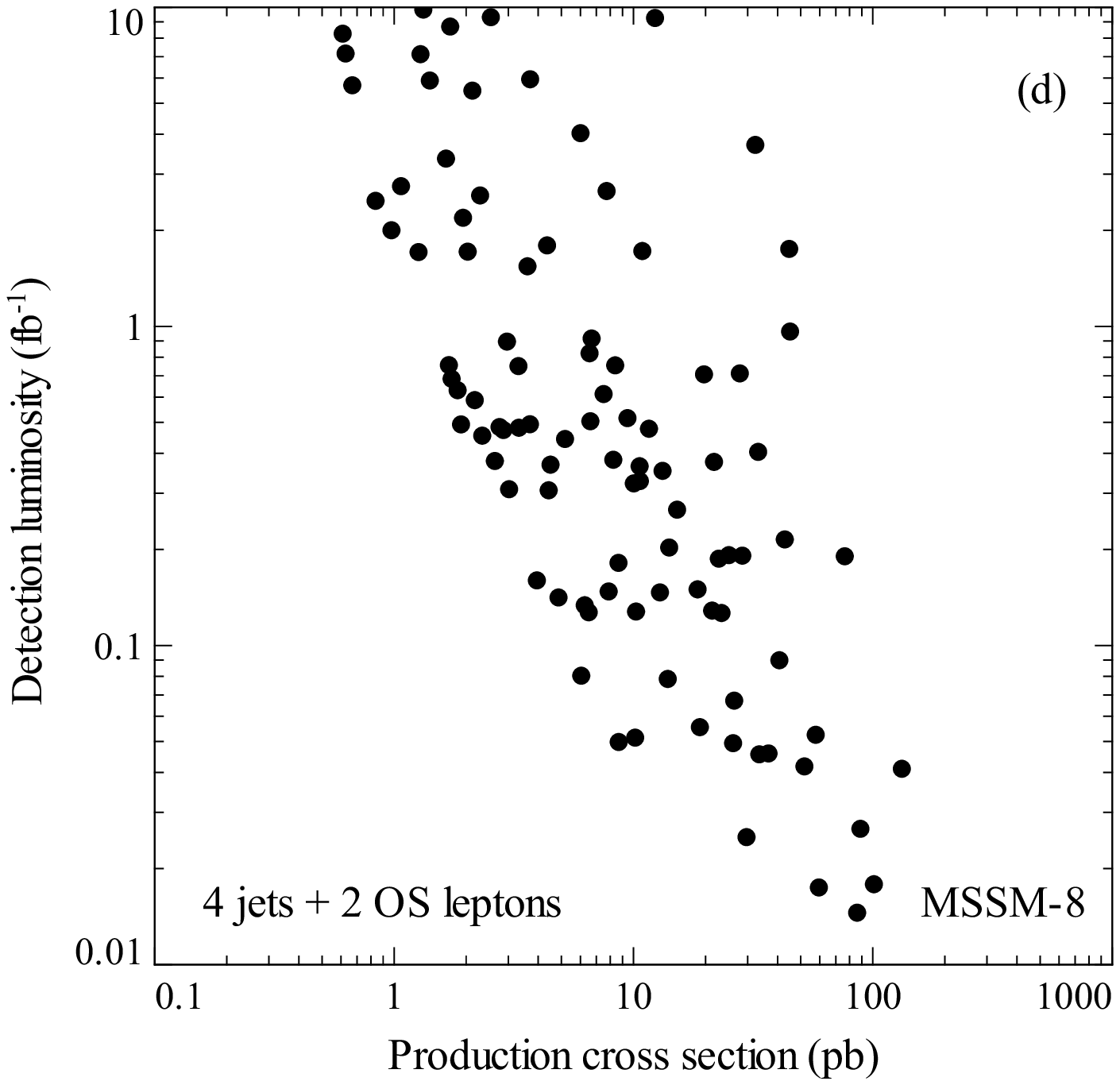}

\includegraphics[scale=0.48]{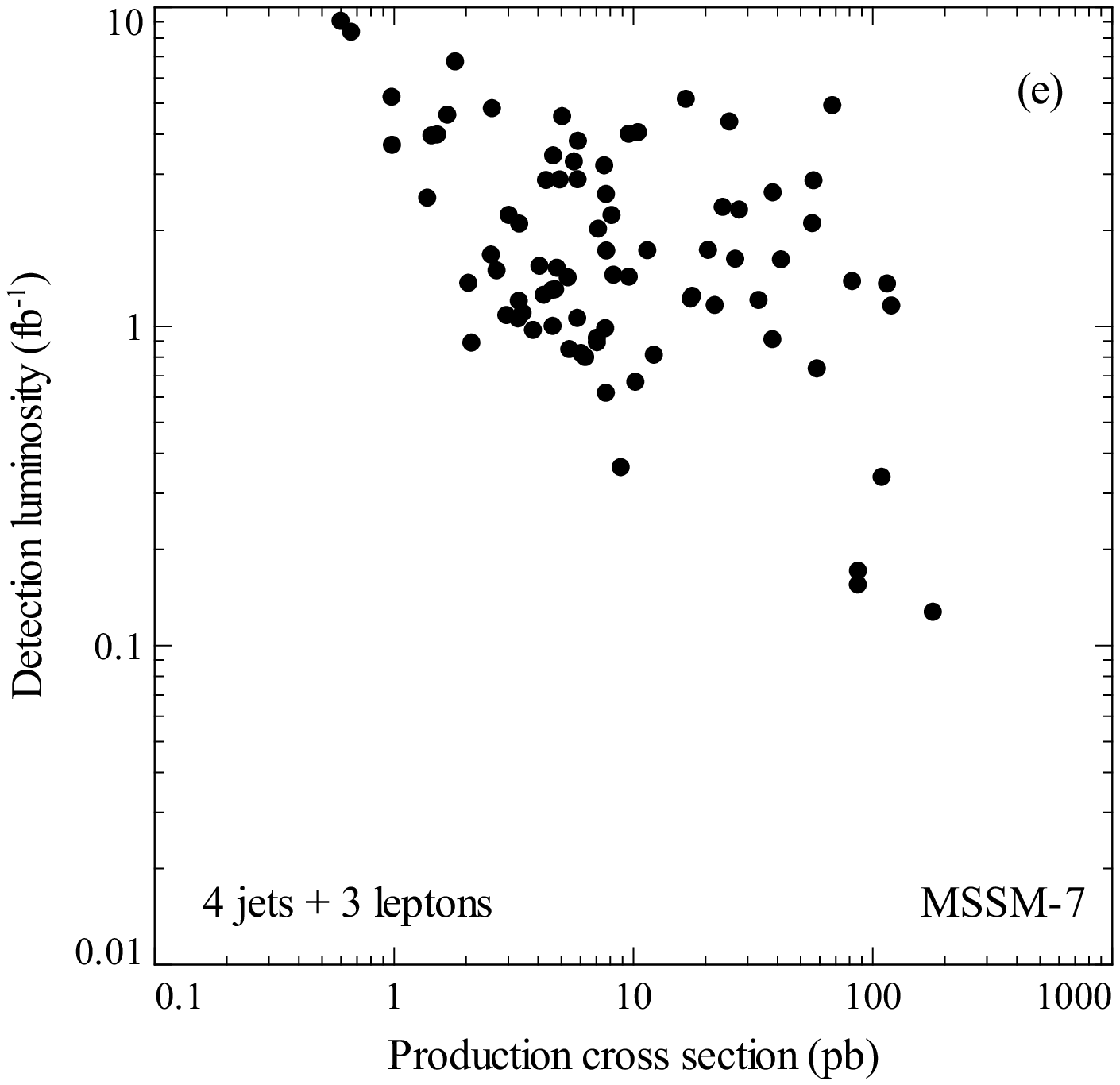}
\includegraphics[scale=0.48]{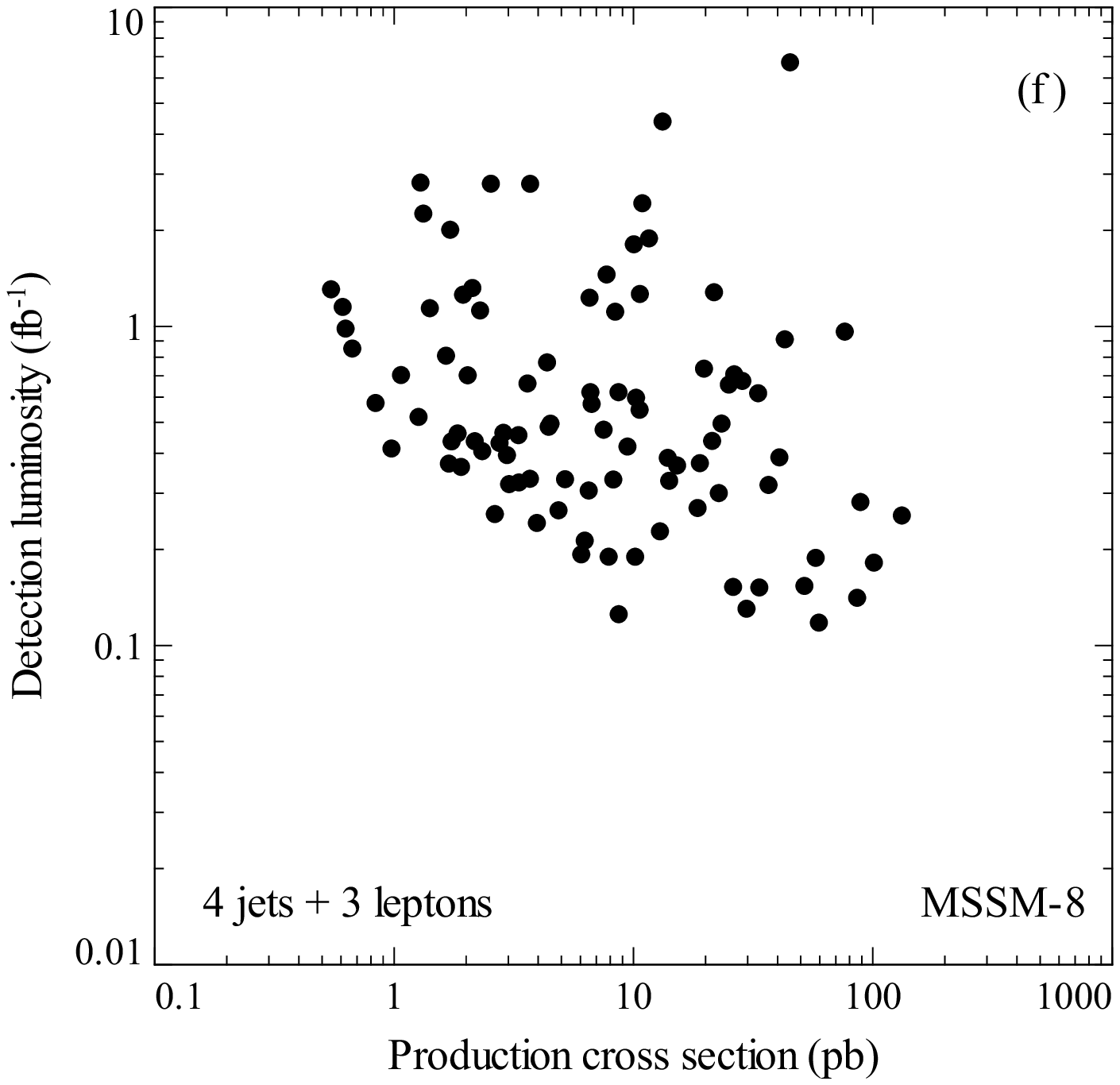}
  \caption{
 Integrated luminosity required for detection {\it versus} total gluino/squark production cross section for the models in our (a),(c),(e) MSSM-7 and (b),(d),(f) MSSM-8 samples. Shown are the results for the (a),(b) SS dilepton, (c),(d) OS dilepton and (e),(f) trilepton search channels.
  }
         \label{fig:corr}
\end{figure}

A summary of the resulting detection luminosities in the four-jet channels can also be found in Figure \ref{fig:corr}, where the integrated luminosity is plotted against the total gluino/squark production cross section. Overall, we find that the prospect for early detection is better for the MSSM-8 models than for the MSSM-7 ones, something which points to the importance of light sleptons in the sparticle decay chain. The correlation between production cross section and detection luminosity is found to be strongest in the OS channel, somewhat weaker in the SS channel and much less prominent in the trilepton channel. Regarding the other model parameters and sparticle mass splittings mentioned in Section \ref{sec:analysis}, no clear-cut correlations with the signal magnitude could be established from our samples.

The channel in which an individual model is first seen varies, although the SS dilepton channel typically looks like the most promising one. The possibility of using the real data to verify the background is however more solid in the OS dilepton channel, which may hence turn out to be of higher interest in upcoming analyses. Indeed the SS dilepton channel should not be much trusted at all before it has been measured in the data (see Section \ref{sec:backgroundproperties}). The results for the trilepton channel, which due to uncertainties in the background estimation should be taken only as rough guidance, are less interesting early on but should come into play as higher luminosities are reached. Often a model shows up in two or more channels, either at about the same time or successively at relatively low luminosities still, offering the possibility to cross-correlate between them.

%==============================================================================
%==============================================================================

\subsection{The four-jet, dilepton channel}
\label{sec:4j2l}
%==============================================================================
\begin{figure}
        \centering
\includegraphics[scale=0.6]{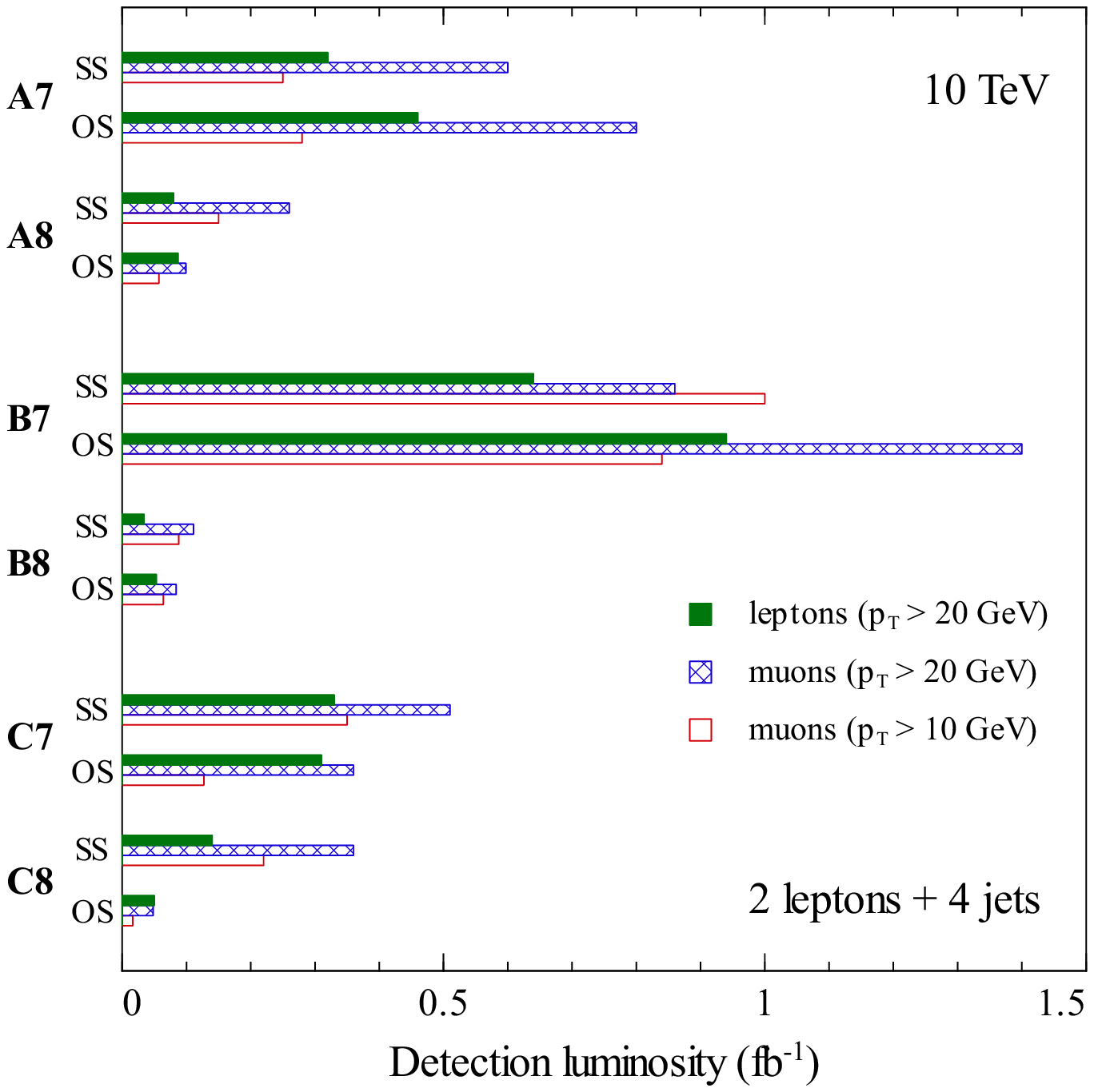}
 \caption{Integrated luminosity required for detection of the benchmark models of Table \ref{table:benchmarkmodels} in the dilepton (electrons+muons) and dimuon channels, at 10 TeV center-of-mass energy. The statistical errors (not shown in the figure) are less than 20\%.}
        \label{fig:bmlum}
\end{figure}
%==============================================================================

\subsubsection{$4 j + 2l$}
\label{subsec:4j+2l}

Figure \ref{fig:4j2l} shows the resulting integrated luminosity required for detection in the four jets plus two leptons channel, to which both electrons and muons with $p_T>20$ GeV contribute. The figure is divided into four parts showing the results for the SS and OS lepton cuts, for our MSSM-7 and MSSM-8 models respectively. Two main observations can be made: the MSSM-8 models have an advantage over the MSSM-7 models in detectability, and the signal typically shows up earlier in the SS channel than in the OS channel.

A majority of the examined MSSM-8 models with gluino masses below $\sim700$ GeV turn out to be detectable within a few hundred pb$^{-1}$ of integrated luminosity. This is true for the signal in both the SS and the OS channel, although the latter typically shows up afterwords, and especially for high squark masses the advantage for early detection within the SS channel is apparent.

Some MSSM-7 models with gluino masses up to $\sim600$ GeV are detectable early on but generally only the region of lighter gluinos and squarks looks promising at low integrated luminosities. With a few exceptions, the SS channel seems to be the most promising one also here.

The advantage of the SS channel may however be fragile. Our calculated background cross section, as given in Table \ref{bgtable}, reveals that a luminosity of the order of 0.1 fb$^{-1}$ must be collected before even a single SS background event is expected. In a real data analysis one would prefer to establish the background level by investigating the response to changes in the imposed cuts, for which a larger number of background events clearly is favorable. This is especially important for the SS background since it consists of lepton fake events (see Section \ref{sec:backgroundproperties}). The OS background, see Table \ref{bgtable}, being an order of magnitude larger and consequently also easier to measure at low integrated luminosities, is on the other hand more robust. Hence the OS channel results may actually be the most important for the early searches.

Another systematic uncertainty in the background determination stems from the fact that, for computational reasons, we include no more than two explicit jets in the hard process generation. This is principally a problem for the background coming from $Z$ production, where ideally at least four explicit jets should be included. For the $t\bar{t}$ background two explicit jets should however give a good accuracy since e.g. $t\rightarrow bW^+$ and $\bar{t}\rightarrow \bar{b}W^-$ decays can give rise to the remaining jets and isolated leptons. As we find the $t\bar{t}$ contribution to the background to clearly dominate the $Z$ contribution within both the SS and OS channel (see Table \ref{bgtable}) we expect this kind of uncertainty to be moderate for the dilepton searches.

\subsubsection{$4j+2\mu$}

\begin{figure}
        \centering
\includegraphics[scale=0.5]{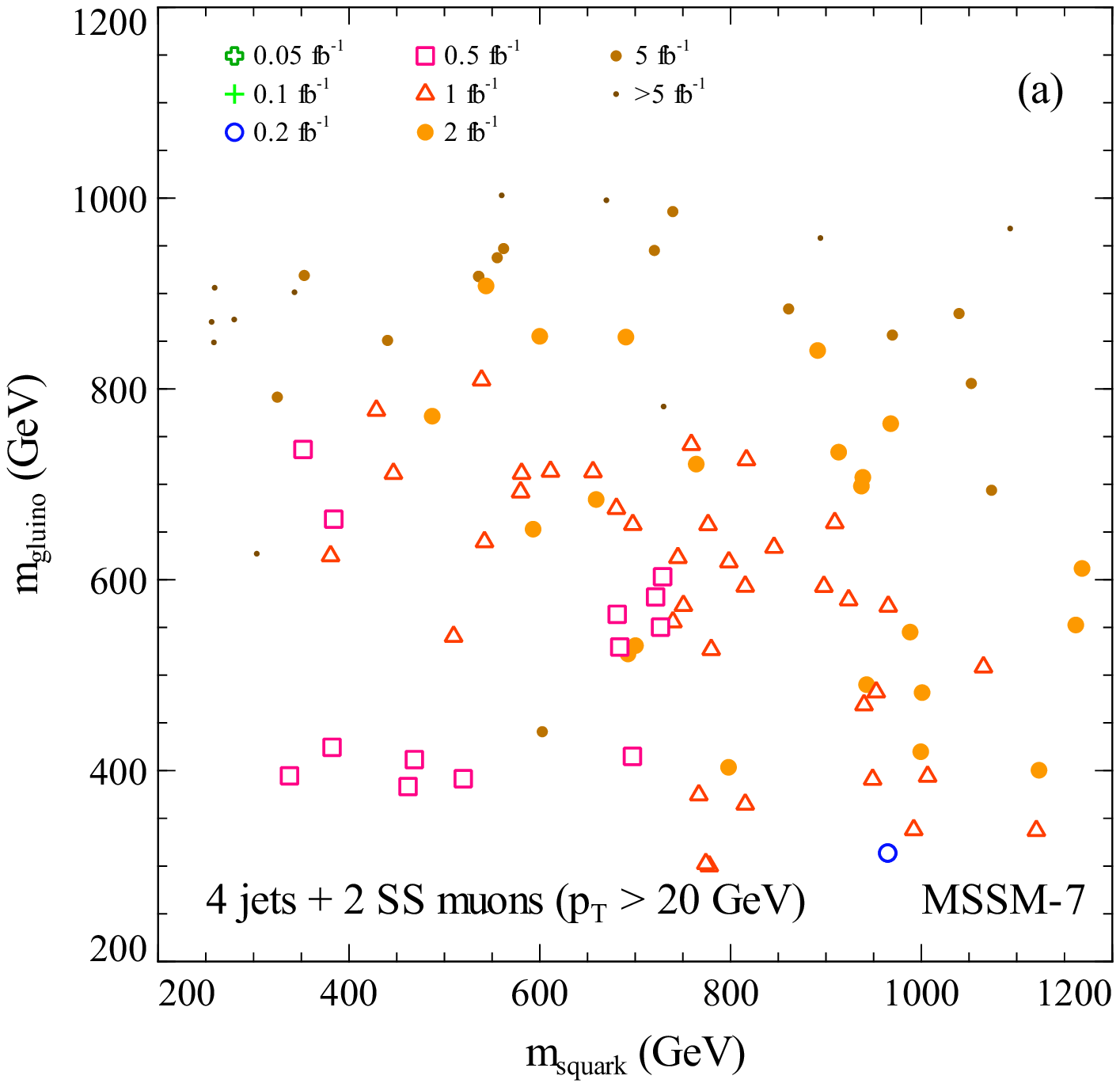}
\includegraphics[scale=0.5]{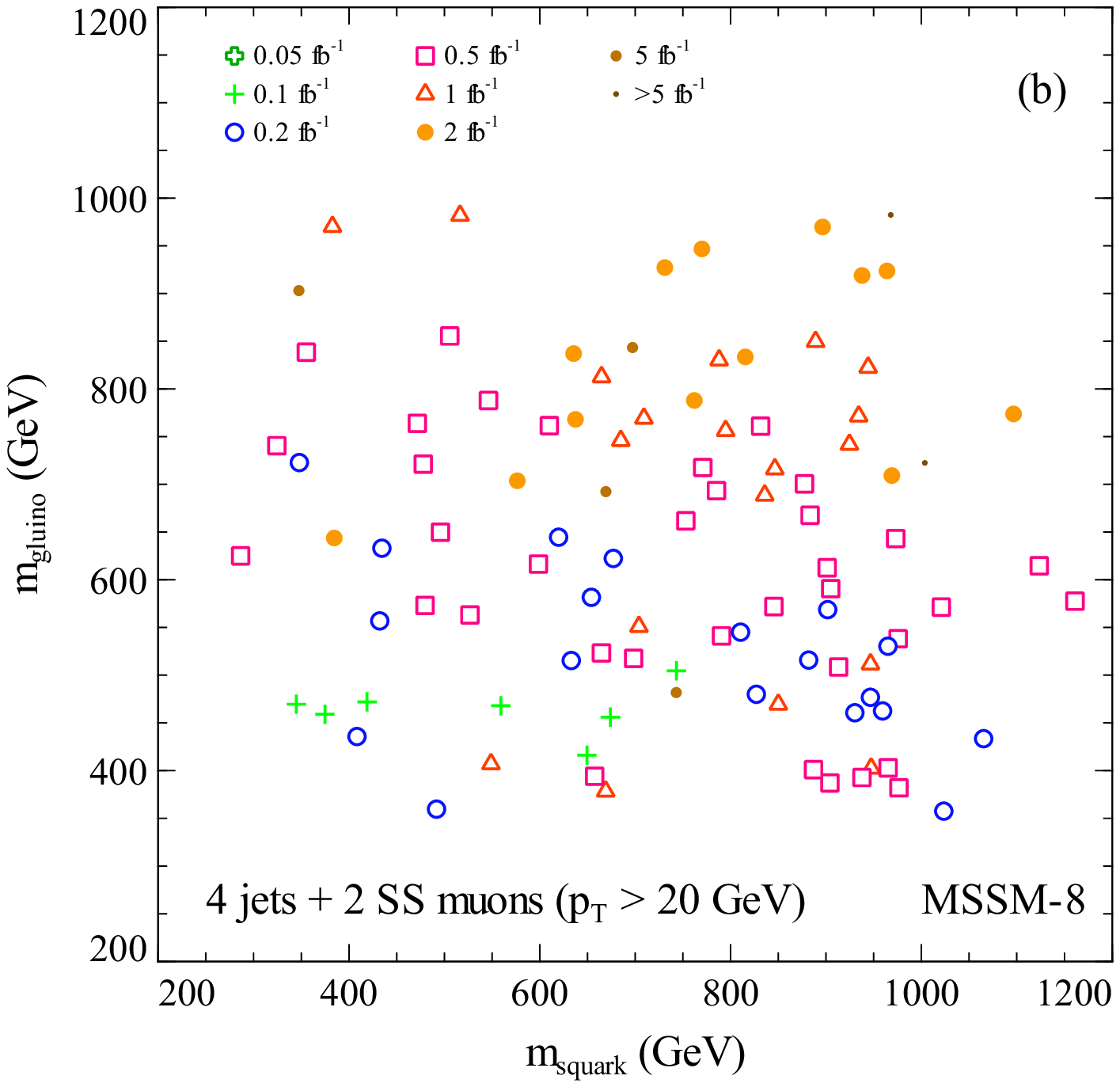}
\includegraphics[scale=0.5]{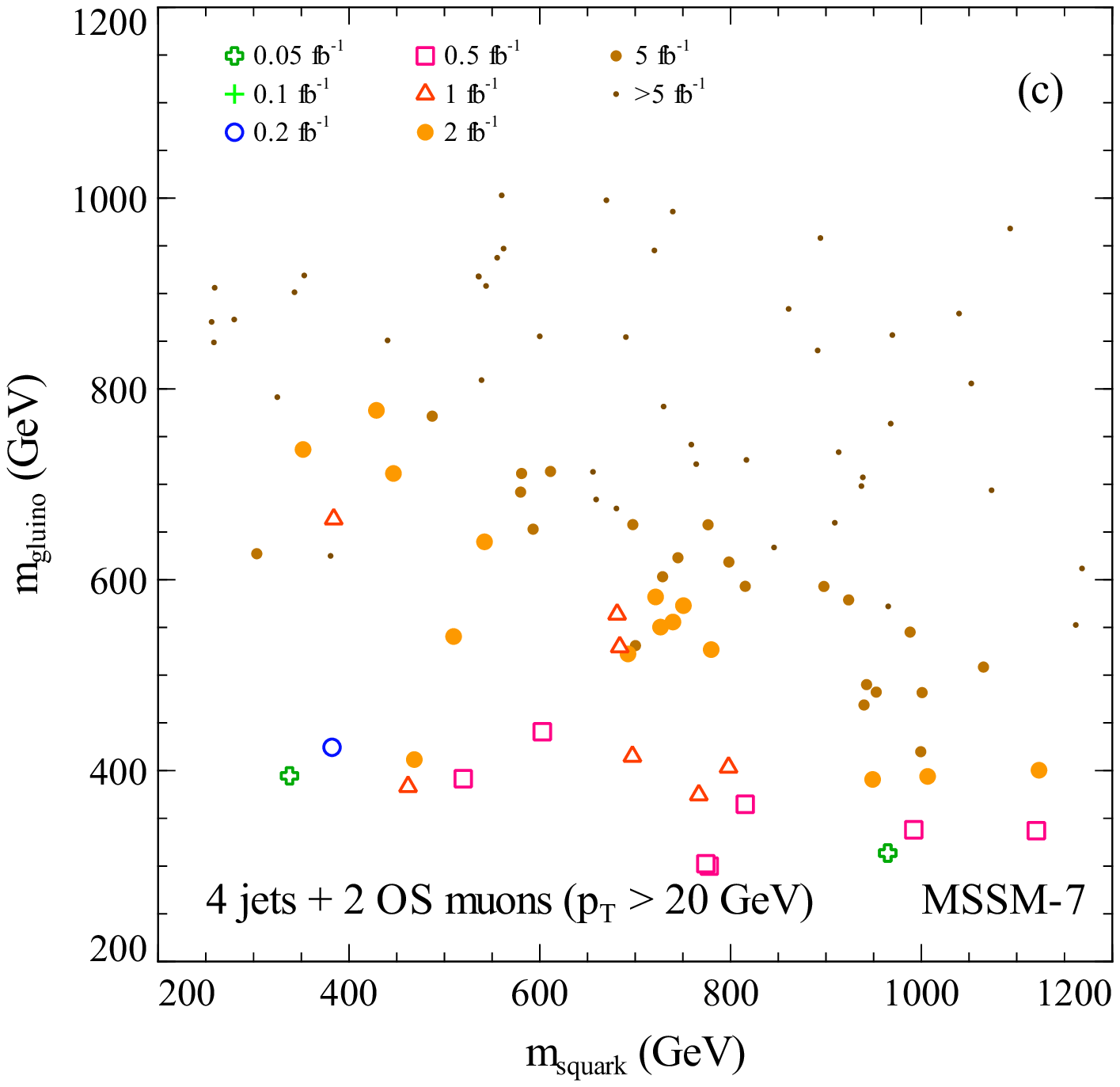}
\includegraphics[scale=0.5]{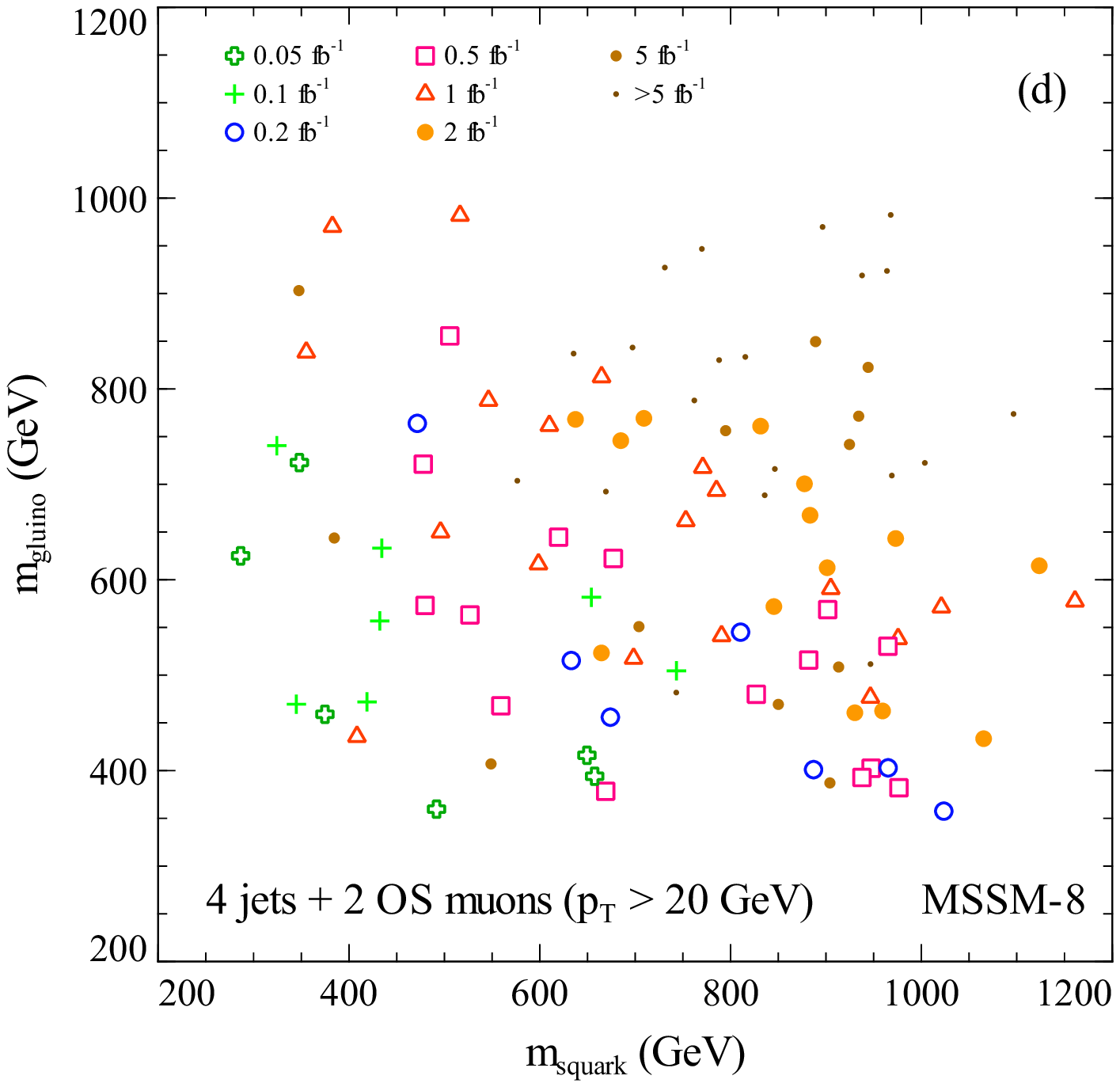}
 	\caption{Integrated luminosity required for detection of (a),(c) MSSM-7 and (b),(d) MSSM-8 models in the four jets (with $p_T>100,50,50,50$ GeV) plus two muons (each with $p_T>20$ GeV) search. The results are shown for the (a),(b) same sign and (c),(d) opposite sign dimuon channels.}
         \label{fig:4j2m20}
\end{figure}

\begin{figure}
        \centering
\includegraphics[scale=0.5]{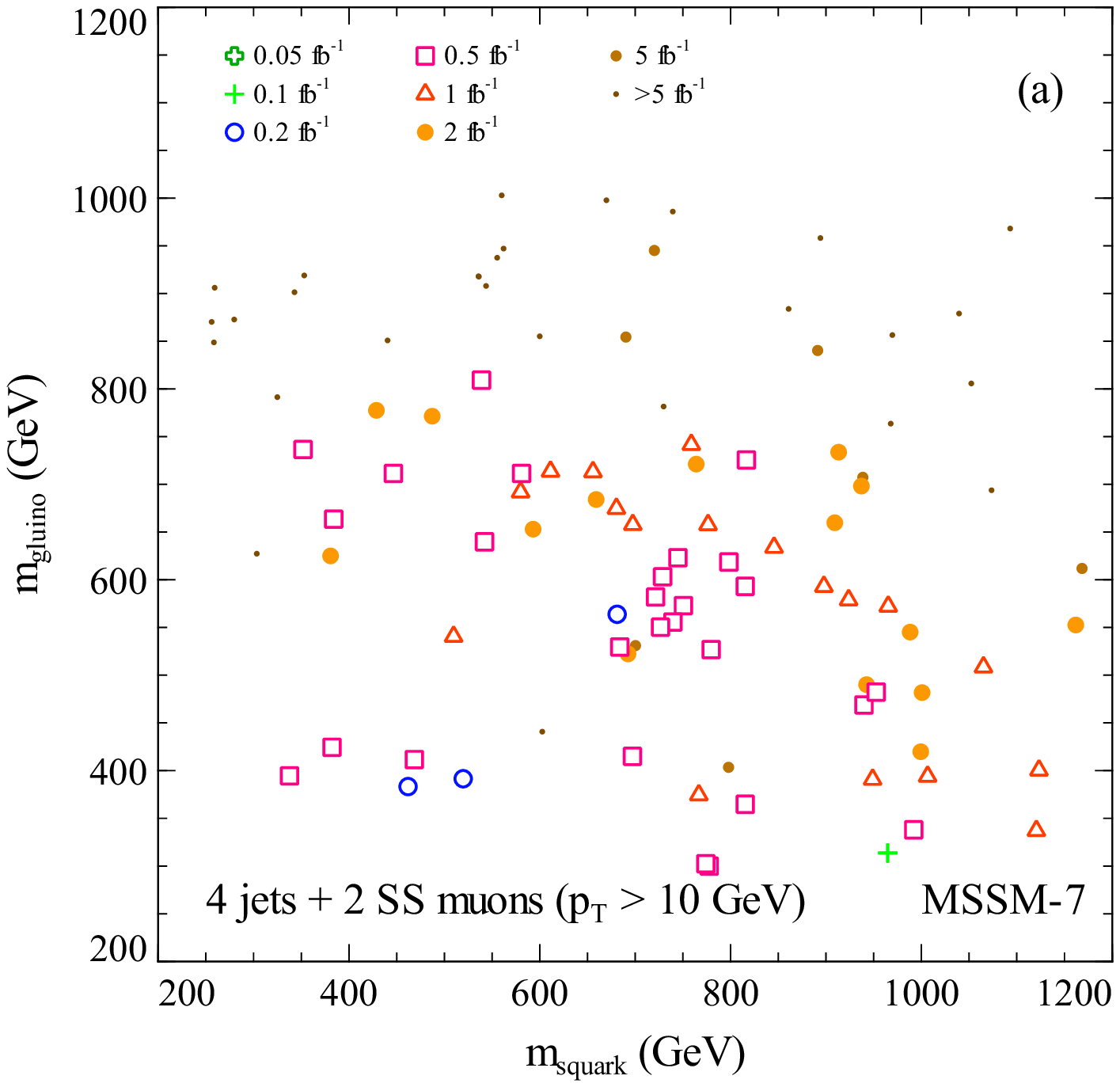}
\includegraphics[scale=0.5]{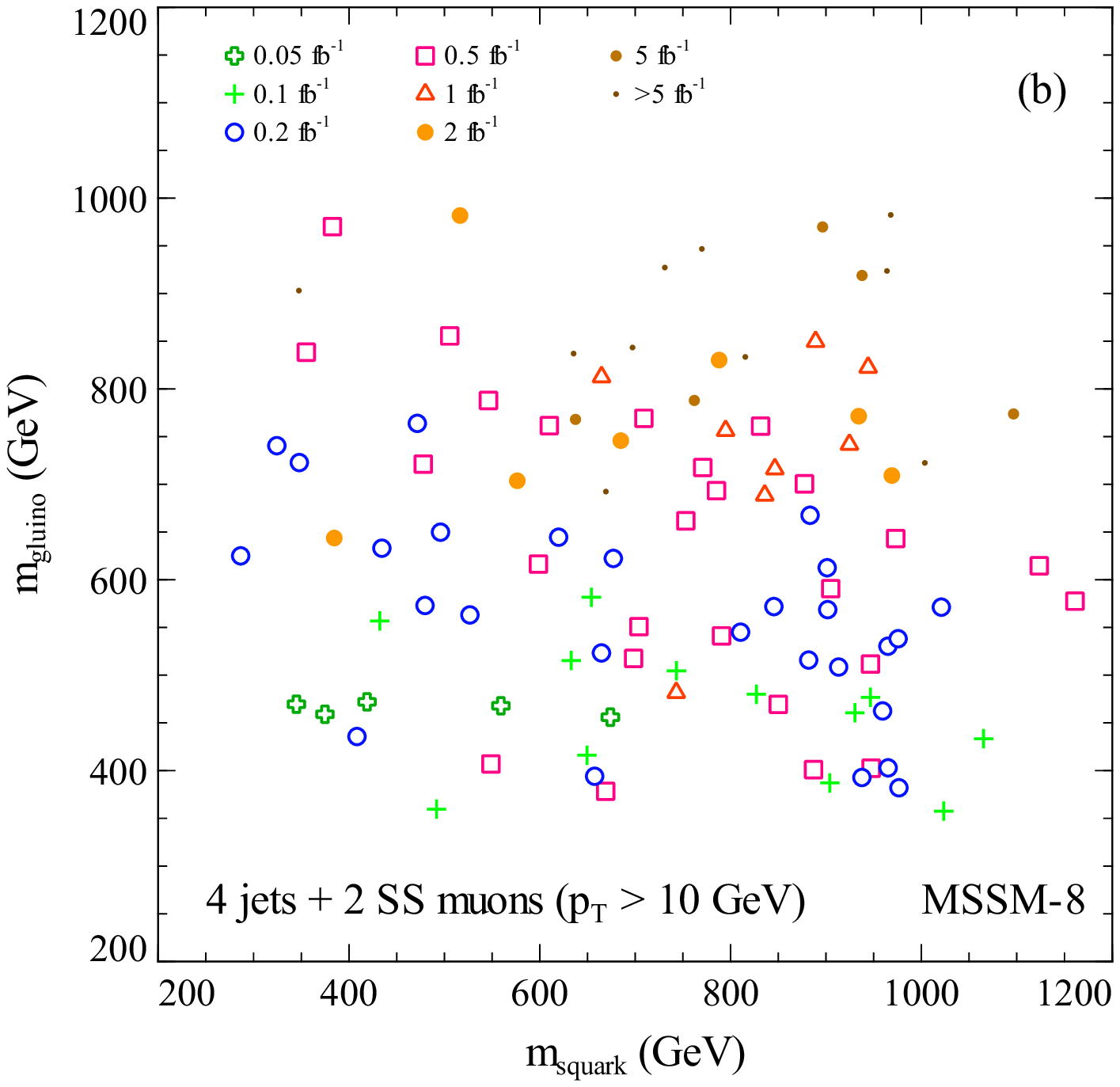}
\includegraphics[scale=0.5]{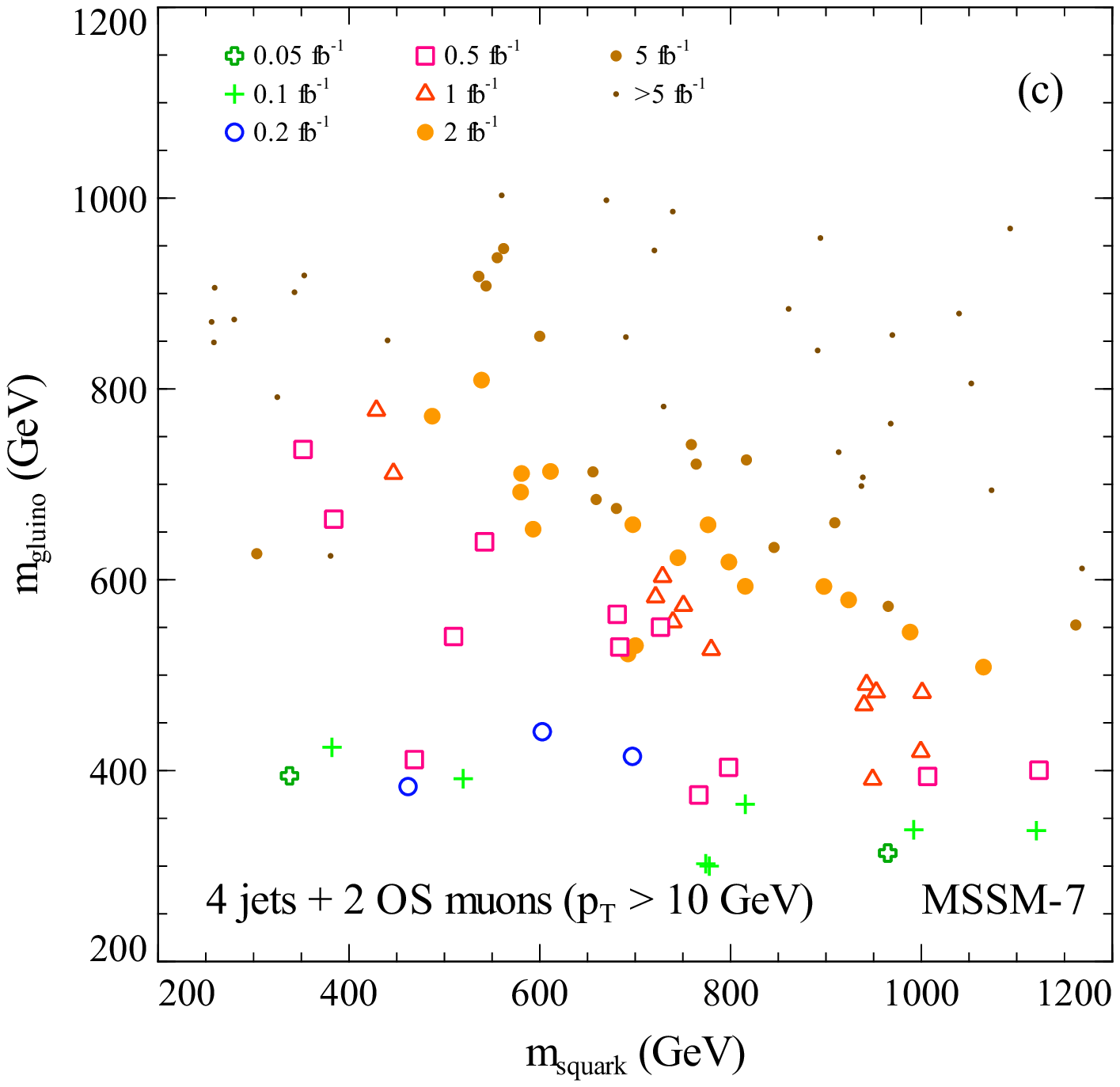}
\includegraphics[scale=0.5]{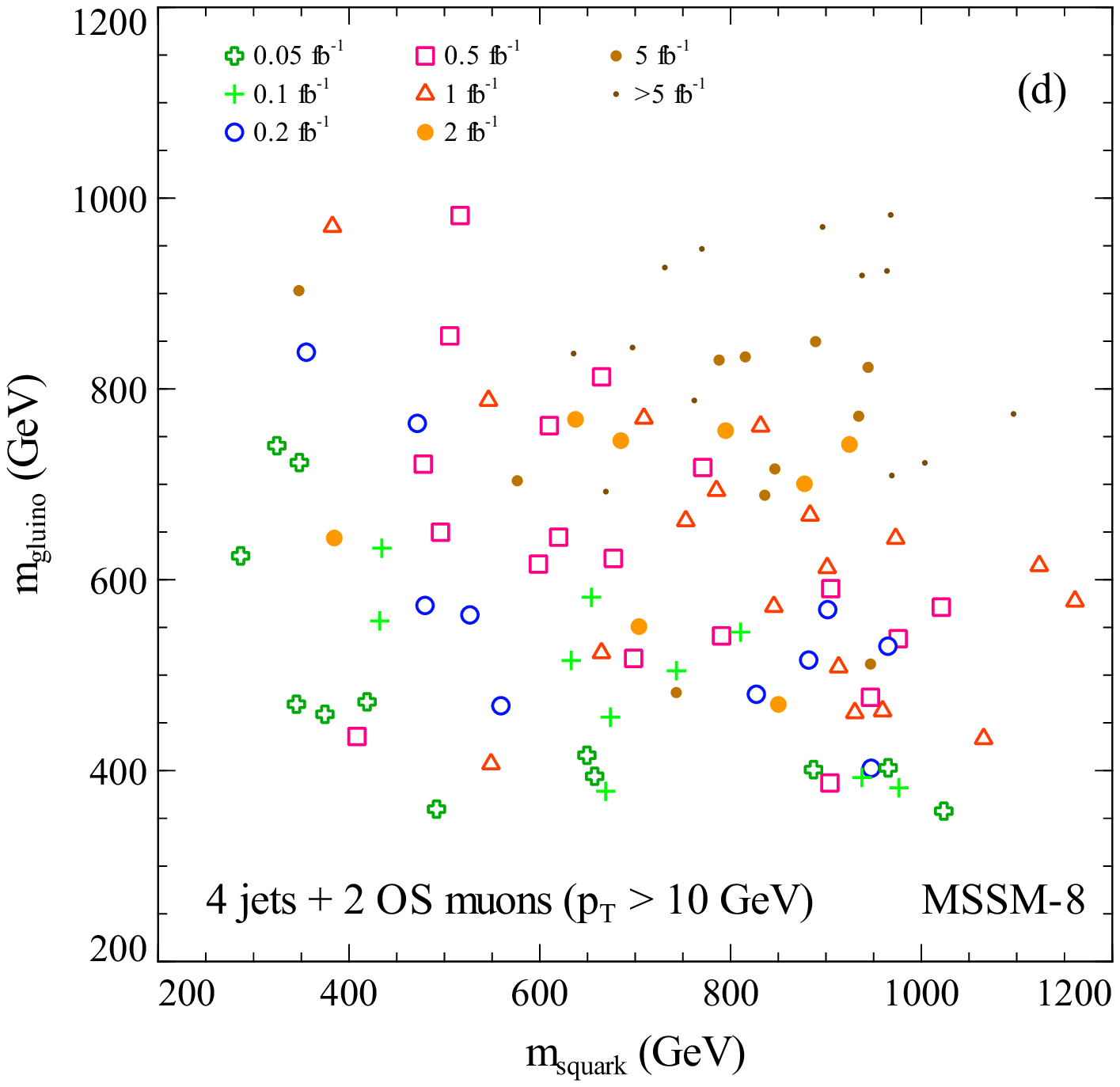}
	\caption{Integrated luminosity required for detection of (a),(c) MSSM-7 and (b),(d) MSSM-8 models in the four jets (with $p_T>100,50,50,50$ GeV) plus two muons (each with $p_T>10$ GeV) search. The results are shown for the (a),(b) same sign and (c),(d) opposite sign dimuon channels.}
        \label{fig:4j2m10}
\end{figure}

In case electron identification does not become reliable until later stages, searches could still focus on muons. Figures \ref{fig:4j2m20}--\ref{fig:4j2m10} show the integrated luminosity required for detection in the four jets plus two muons channels. Following the discussion in Section \ref{sec:searchchannels} we show the results for two different cuts on the muon $p_T$.

Figure \ref{fig:4j2m20} shows the resulting detection luminosities for the dimuon search using the $p_T>20$ GeV requirement. Overall the detection prospects are not as good as when electrons are included, although many MSSM-8 models with gluinos lighter than $\sim600$ GeV can still be found within the early data. Whether the SS or the OS channel is the most advantageous varies from model to model. The expected background in Table \ref{bgtable} indicates that several fb$^{-1}$ must be collected before the first SS dimuon $t\bar{t}$ event shows up, which means that a real data verification of that background will hardly be feasible and the most solid early time results will probably rely on an OS dimuon analysis.

In Figure \ref{fig:4j2m10} the corresponding results for the $p_T>10$ GeV muon cut can be found. For basically every model, detection can happen sooner than for the higher muon $p_T$ requirement. For the OS channel the prospects are often even better than for the $p_T>20$ GeV dilepton search. Again the relative strength of the SS and OS signals varies, and after a couple of hundreds of pb$^{-1}$ many models with gluino masses up to $\sim400$ GeV (MSSM-7) or $\sim600$ GeV (MSSM-8) should be clearly detectable.

\subsubsection{Benchmark models}

In Figure \ref{fig:bmlum} we show the resulting integrated luminosities needed for detection of our benchmark models, defined in Table \ref{table:benchmarkmodels}, in the four jets plus two leptons searches that we have considered. As the two models within each pair only differ in the slepton masses, the figure cleary illustrates the advantage of having sleptons accessible in the sparticle decay chain. Note also that the relative significance of the various search channels varies between the models.

\subsection{The four-jet, trilepton channel}
\label{sec:4j3l}

As discussed in Section \ref{sec:4j2l} the magnitude of especially the $Z$ background comes with some uncertainty due to practical constraints on the number of explicit hard jets feasible to include in our simulation. Unlike what we find for the dilepton channels, $Z$ production seems to dominate the background for the trilepton searches, see Table \ref{bgtable}. This means that the systematic uncertainty may be rather severe within this channel. In addition, the sample size needed to find an accurate expected background is rather large, as can be seen from the statistical errors due to our simulation indicated in Table \ref{bgtable}.

Nevertheless, we have checked that our resulting background agrees fairly well with that found in \cite{Baer2009}, in particular in that the five-event level of Eq.~(\ref{eq:detectionlimit}) is what determines the detection up to fairly high integrated luminosities: $\sim1$ fb$^{-1}$ for $p_T>20$ GeV leptons and $\sim10$ fb$^{-1}$ for $p_T>10$ GeV muons in our case, see Table \ref{bgtable}.

\begin{figure}
        \centering
\includegraphics[scale=0.5]{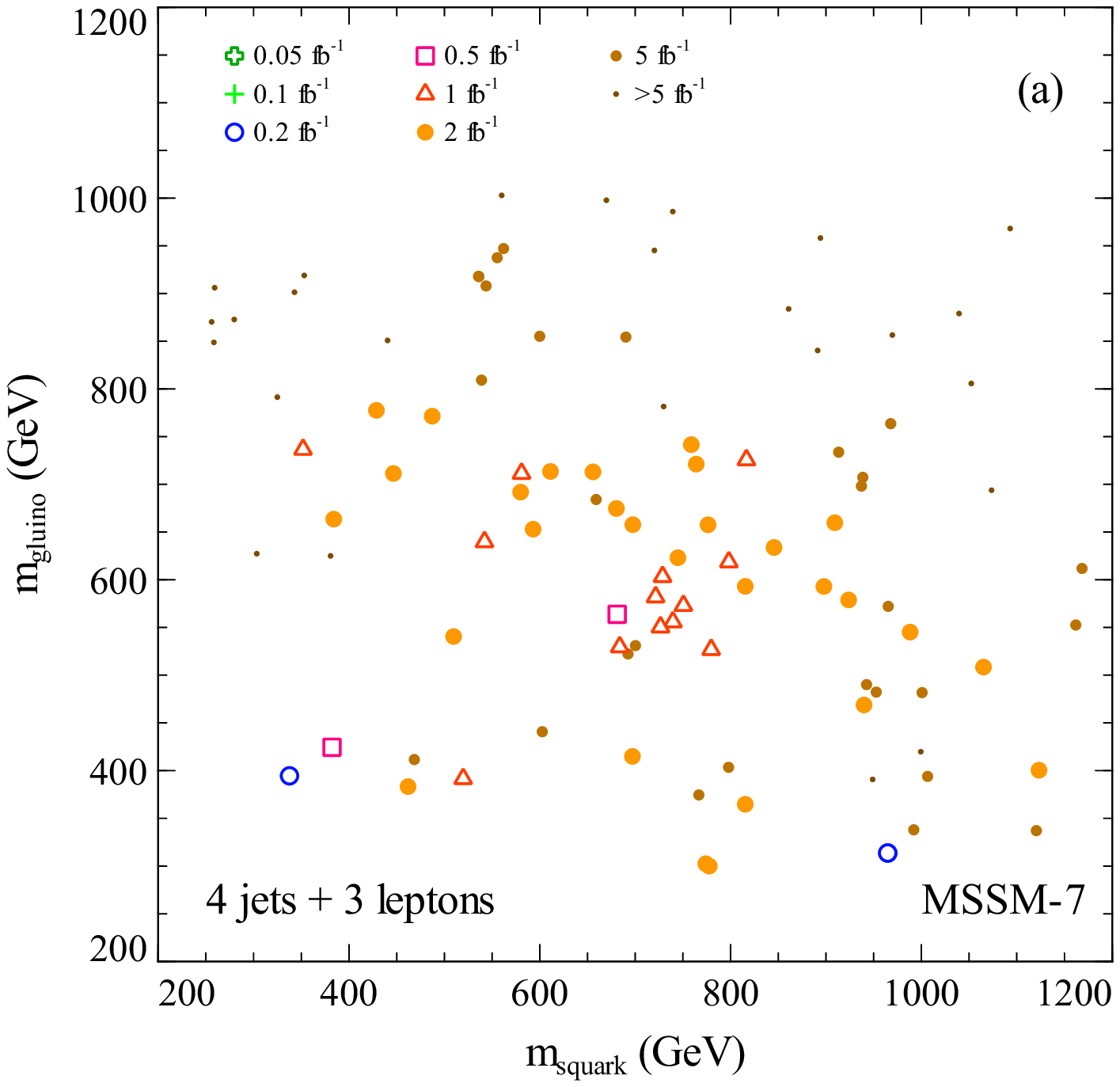}
\includegraphics[scale=0.5]{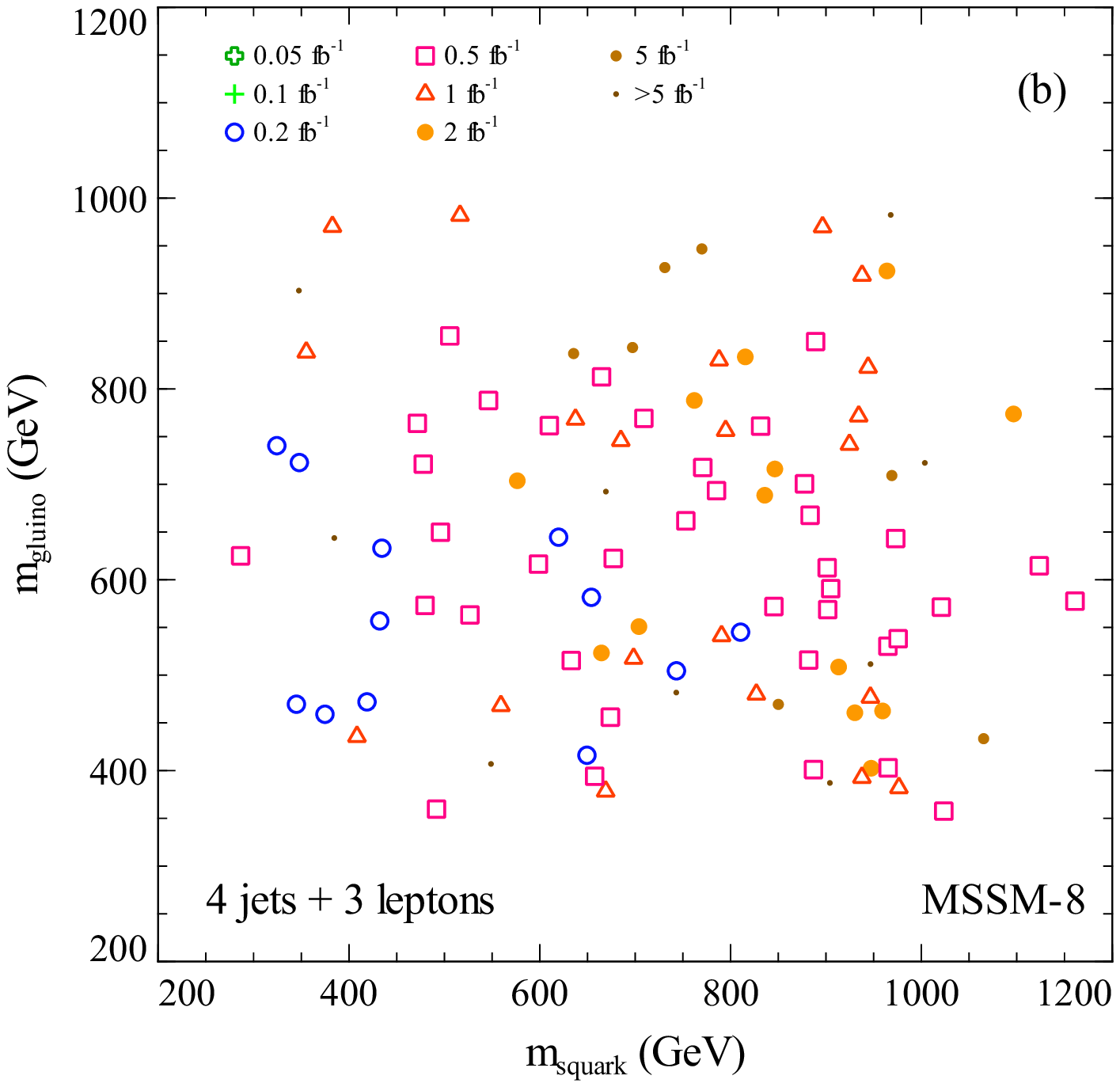}
	\caption{Integrated luminosity required for detection of (a) MSSM-7 and (b) MSSM-8 models in the four jets (with $p_T>100,50,50,50$ GeV) plus three leptons (each with $p_T>20$ GeV) search.}
        \label{fig:4j3l}
\end{figure}

\begin{figure}
        \centering
\includegraphics[scale=0.5]{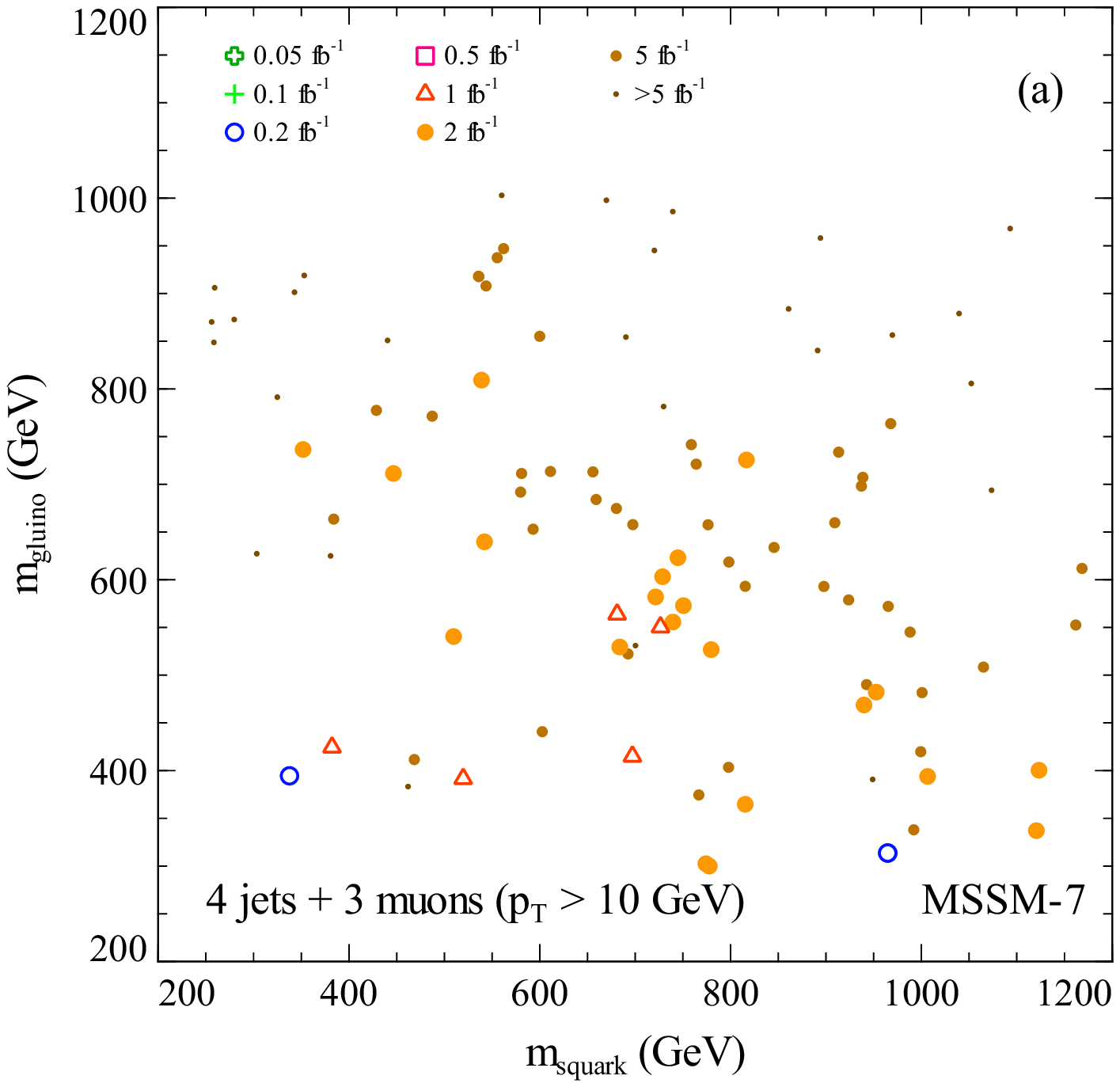}
\includegraphics[scale=0.5]{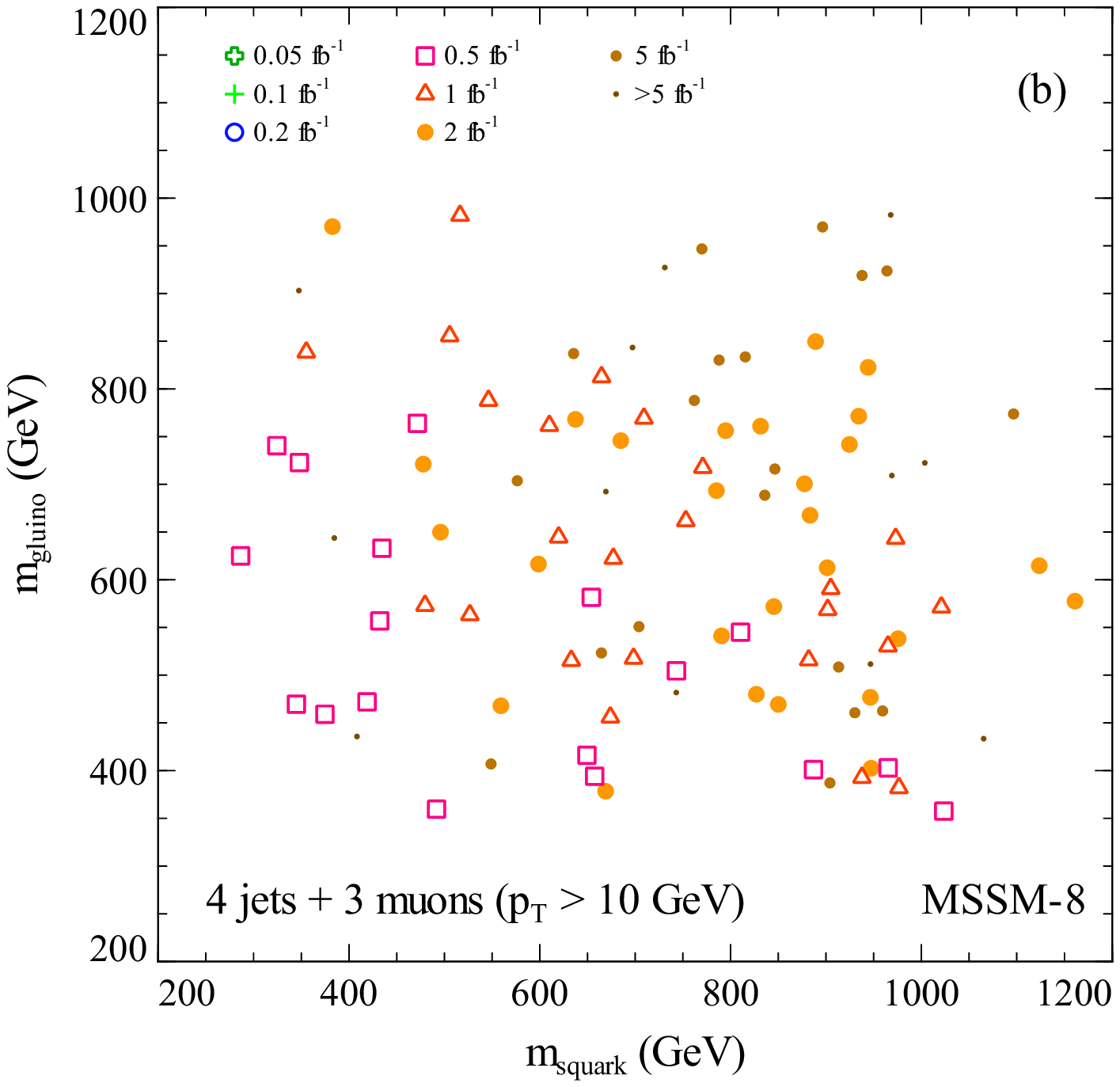}
	\caption{Integrated luminosity required for detection of (a) MSSM-7 and (b) MSSM-8 models in the four jets (with $p_T>100,50,50,50$ GeV) plus three muons (each with $p_T>10$ GeV) search.}
        \label{fig:4j3m10}
\end{figure}

We therefore, despite the uncertainty in the background determination, show plots of our resulting LHC reach within the $p_T>20$ GeV trilepton (electrons and muons) and the $p_T>10$ GeV trimuon channels in Figures \ref{fig:4j3l}--\ref{fig:4j3m10}. While detection of some MSSM-8 models seems to be achievable before $200$ pb$^{-1}$, the prospects are generally worse than for the dilepton channels, a conclusion which agrees with that of \cite{Baer2008,Baer2009}. The trilepton channels could be used later on to pin down the properties of the model underlying a detected signal. 

\subsection{The three-jet channels}
\label{sec:3jl}

\begin{figure}
        \centering
\includegraphics[scale=0.5]{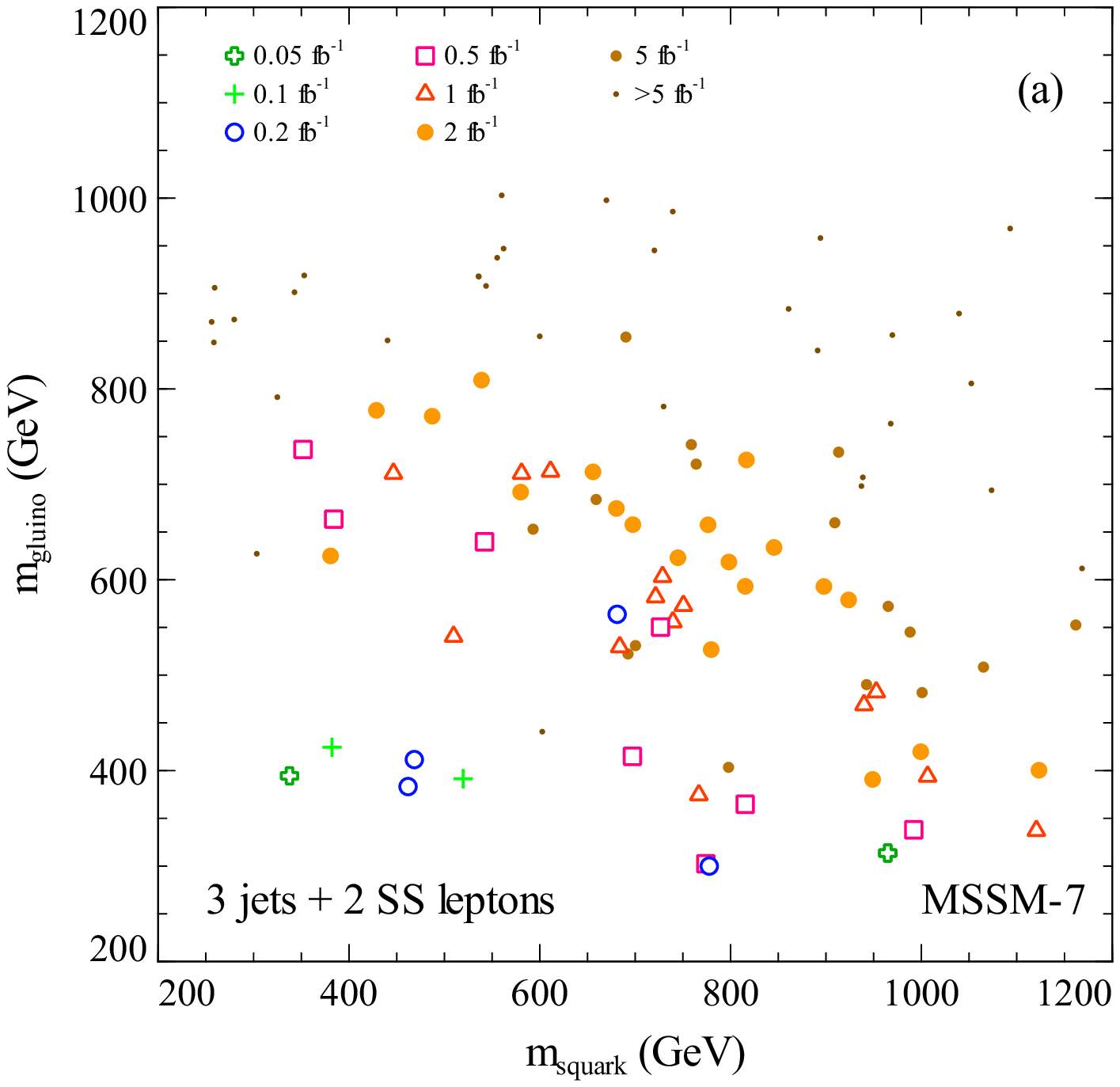}
\includegraphics[scale=0.5]{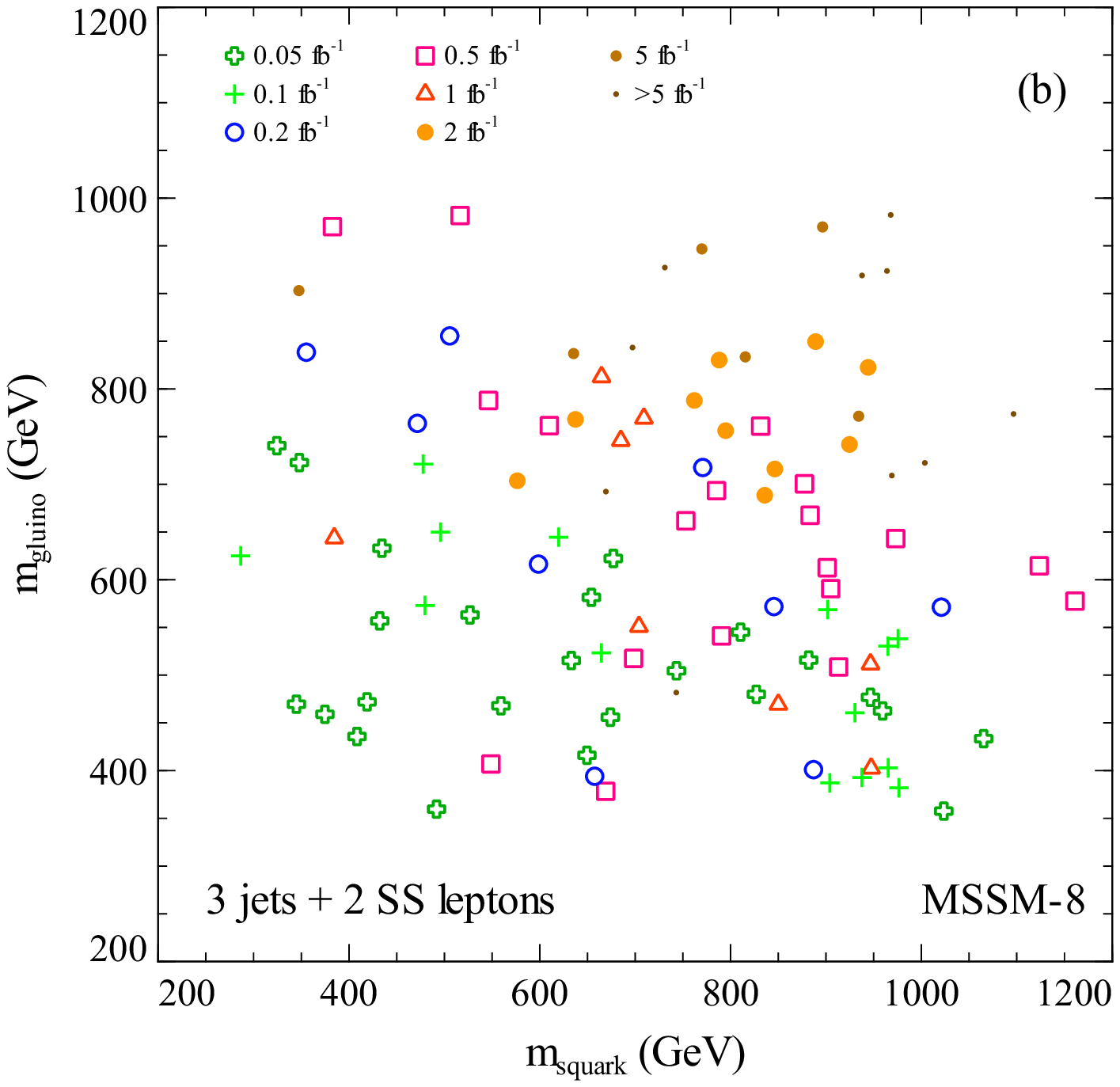}
\includegraphics[scale=0.5]{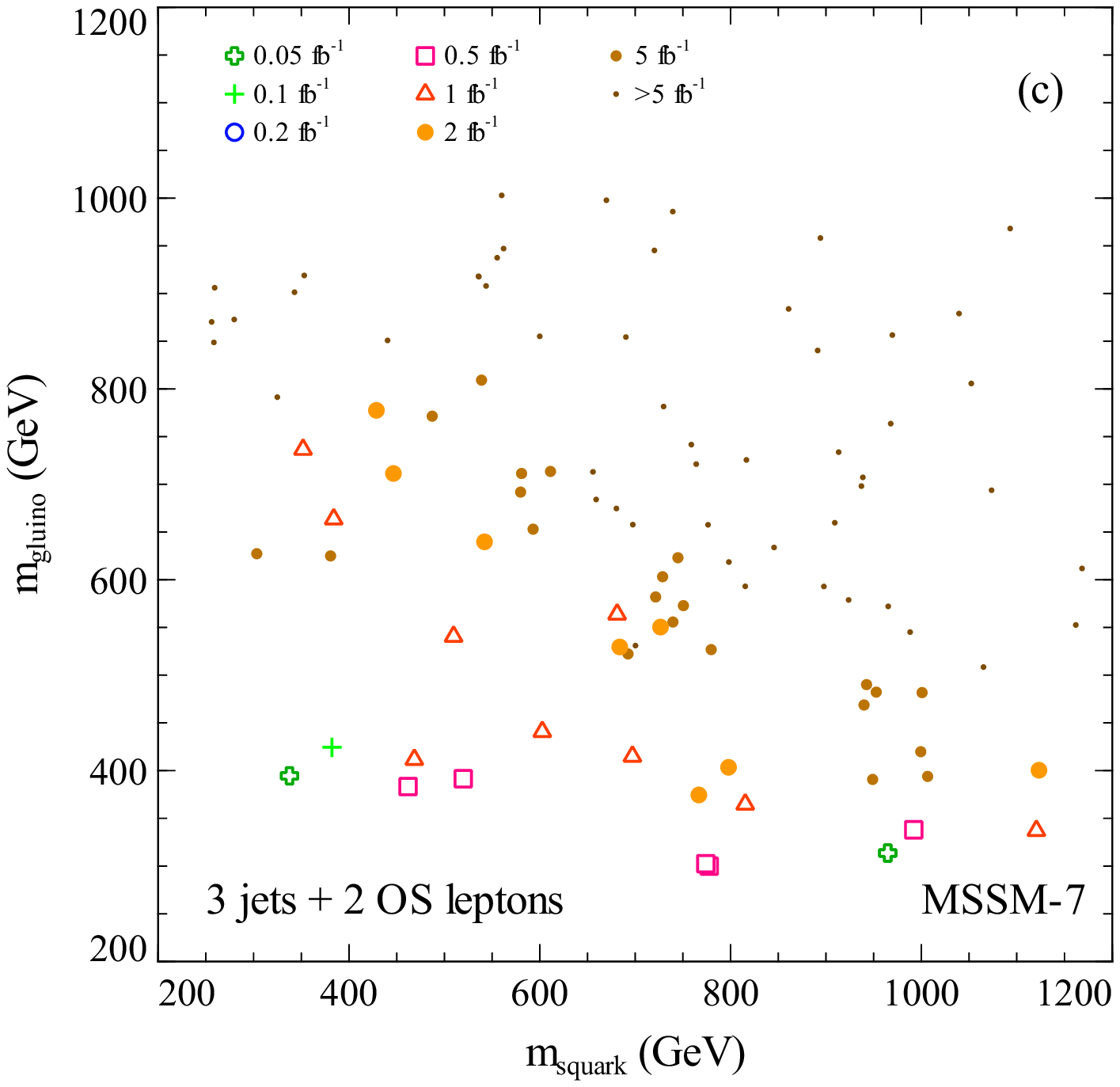}
\includegraphics[scale=0.5]{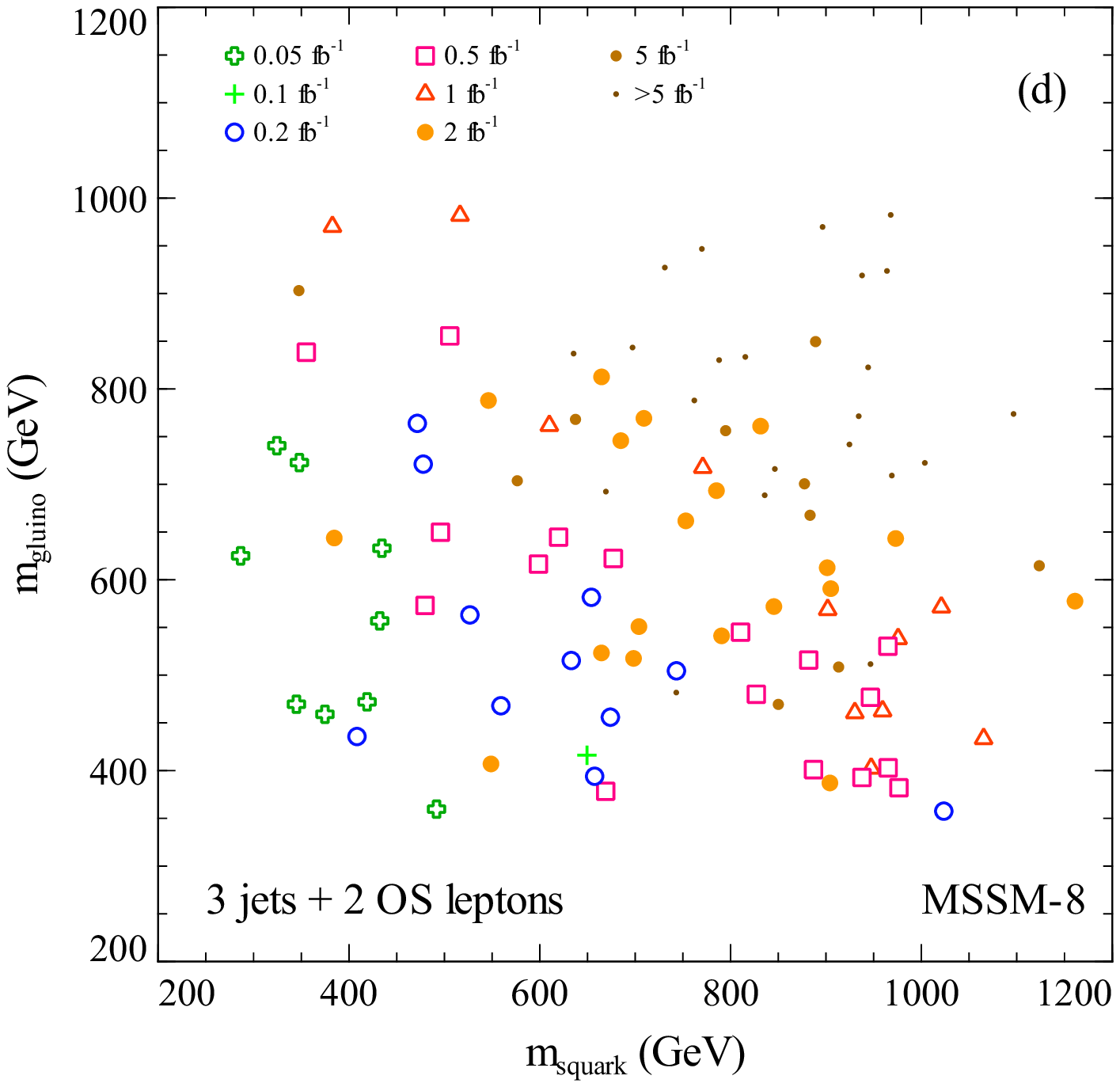}
	\caption{Integrated luminosity required for detection of (a),(c) MSSM-7 and (b),(d) MSSM-8 models in the three jets (with $p_T>50,50,50$ GeV) plus two leptons (each with $p_T>20$ GeV) search. The results are shown for the (a),(b) same sign and (c),(d) opposite sign dilepton channels.}
        \label{fig:3j2l}
\end{figure}

Assuming the same processes to contribute to the background as in the four-jet searches, we show the results for the three jets plus two leptons channel in Figure \ref{fig:3j2l}. A comparison with the corresponding four-jet results of Figure \ref{fig:4j2l} indicates that the four-jet channel seems slightly more favorable, especially in the OS dilepton channel and for heavy gluinos and squarks.

Basically, the same properties hold for the three-jet plus dimuon search, while the three-jet plus trilepton search suffers from large systematic errors within the simulated standard model background. Neither of those results are presented here.

\subsection{Systematic uncertainties}
\label{sec:systematicuncertainties}
%==============================================================================
\begin{figure}
        \centering
\includegraphics[scale=0.32]{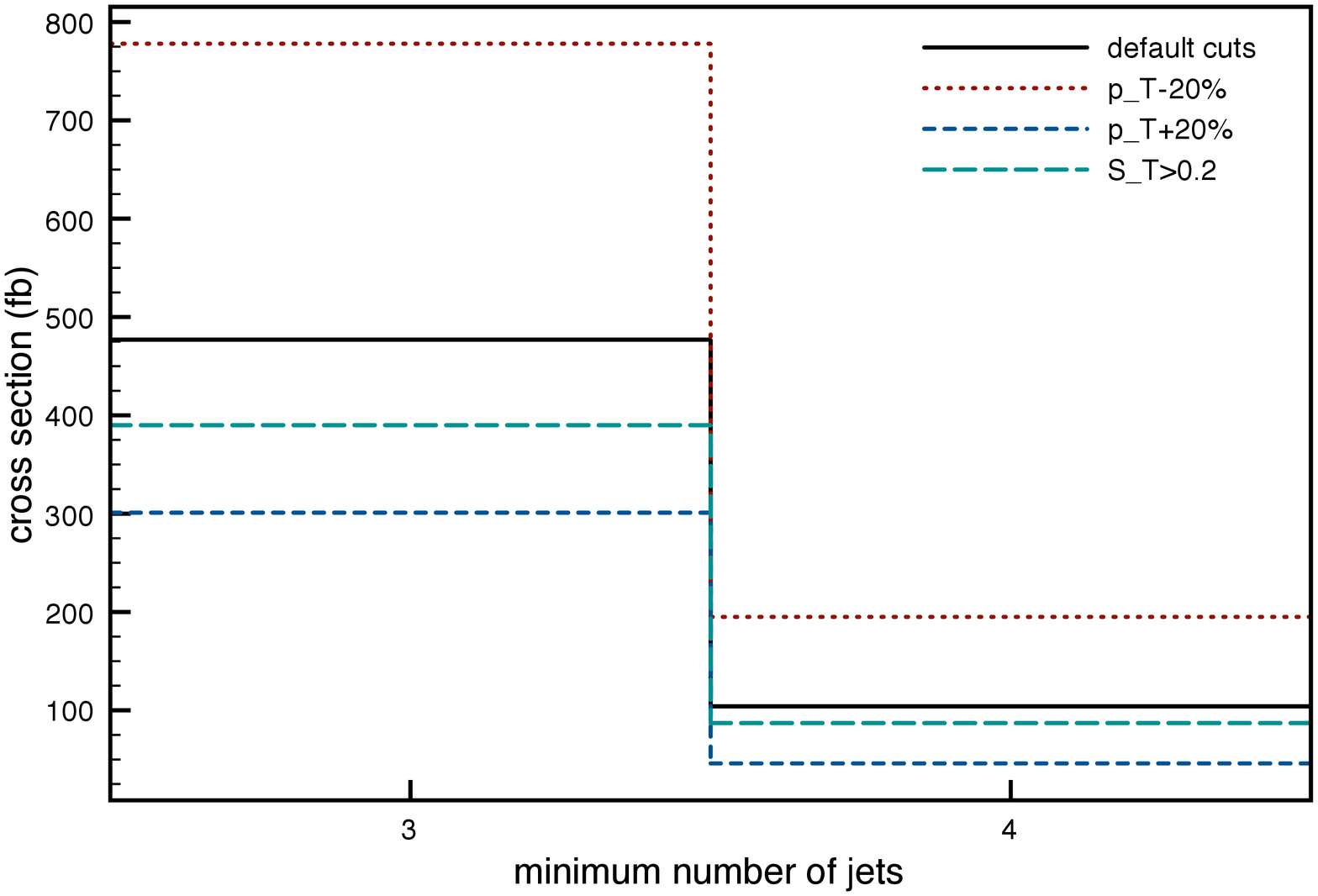}
\includegraphics[scale=0.32]{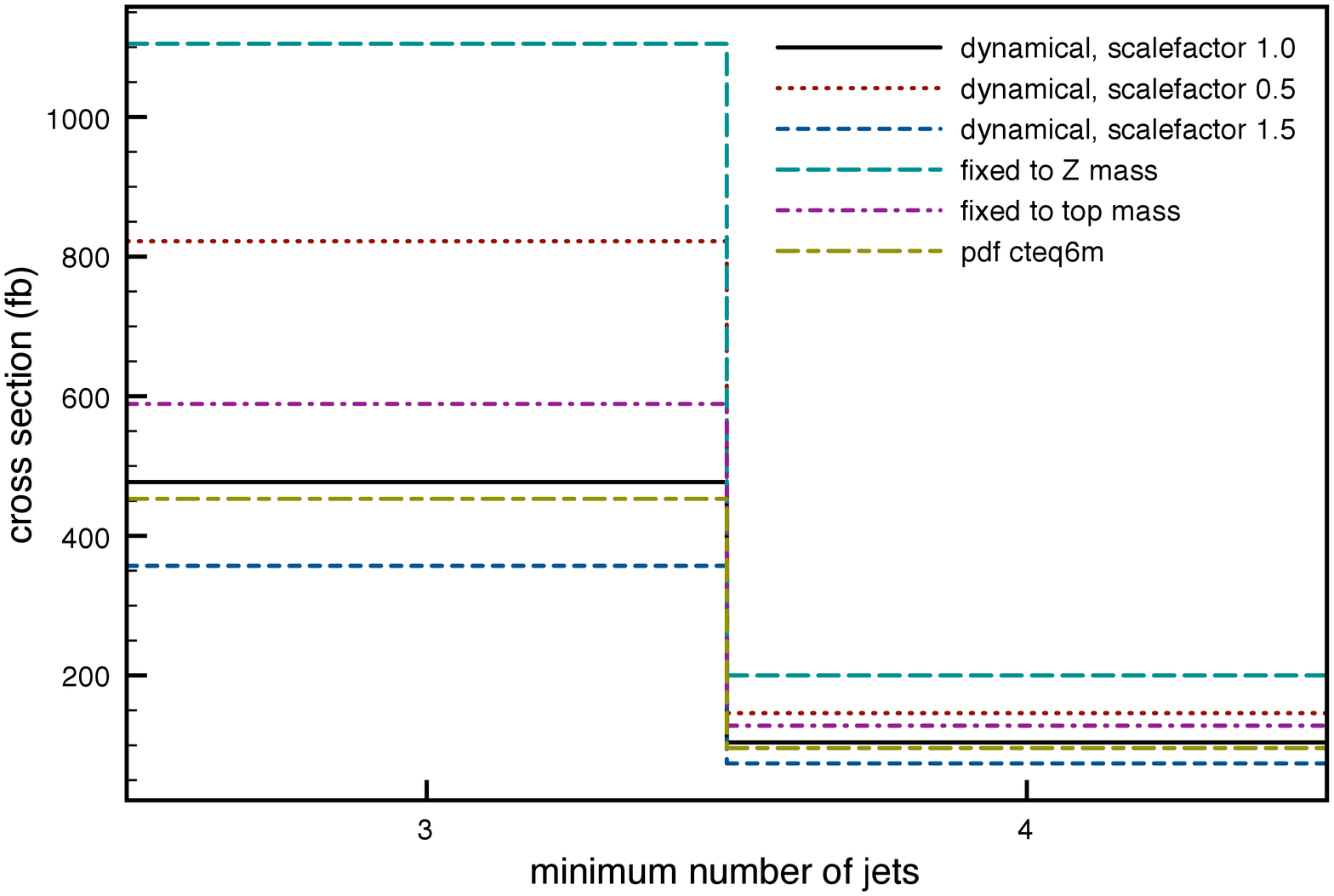}
 \caption{Tests of jet systematics on the $t\bar{t}$ background in the dilepton channel. On the left we show the effect of a 20\% uncertainty in the jet energy determination in the early data, and that the effect of a sphericity cut is small in comparison. On the right we show the effect of different choices for the renormalization and factorization scales as well as the effect of using a different pdf. The solid black line corresponds to our analysis. The \emph{scalefactor} is an invoked rescaling factor common for all events.}
        \label{fig:syst1}
\end{figure}
%==============================================================================
%==============================================================================
\begin{figure}
        \centering
\includegraphics[scale=0.32]{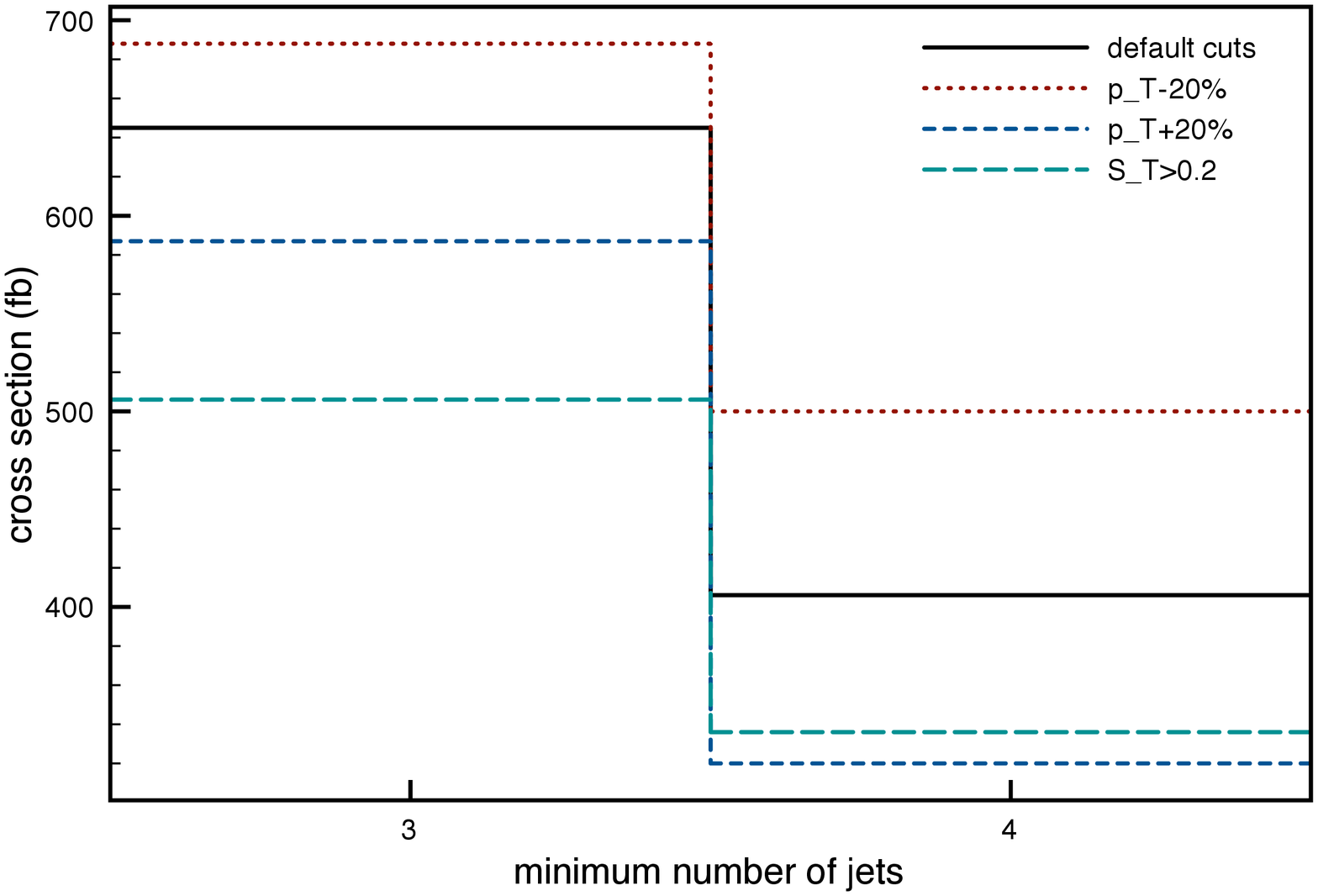}
\includegraphics[scale=0.32]{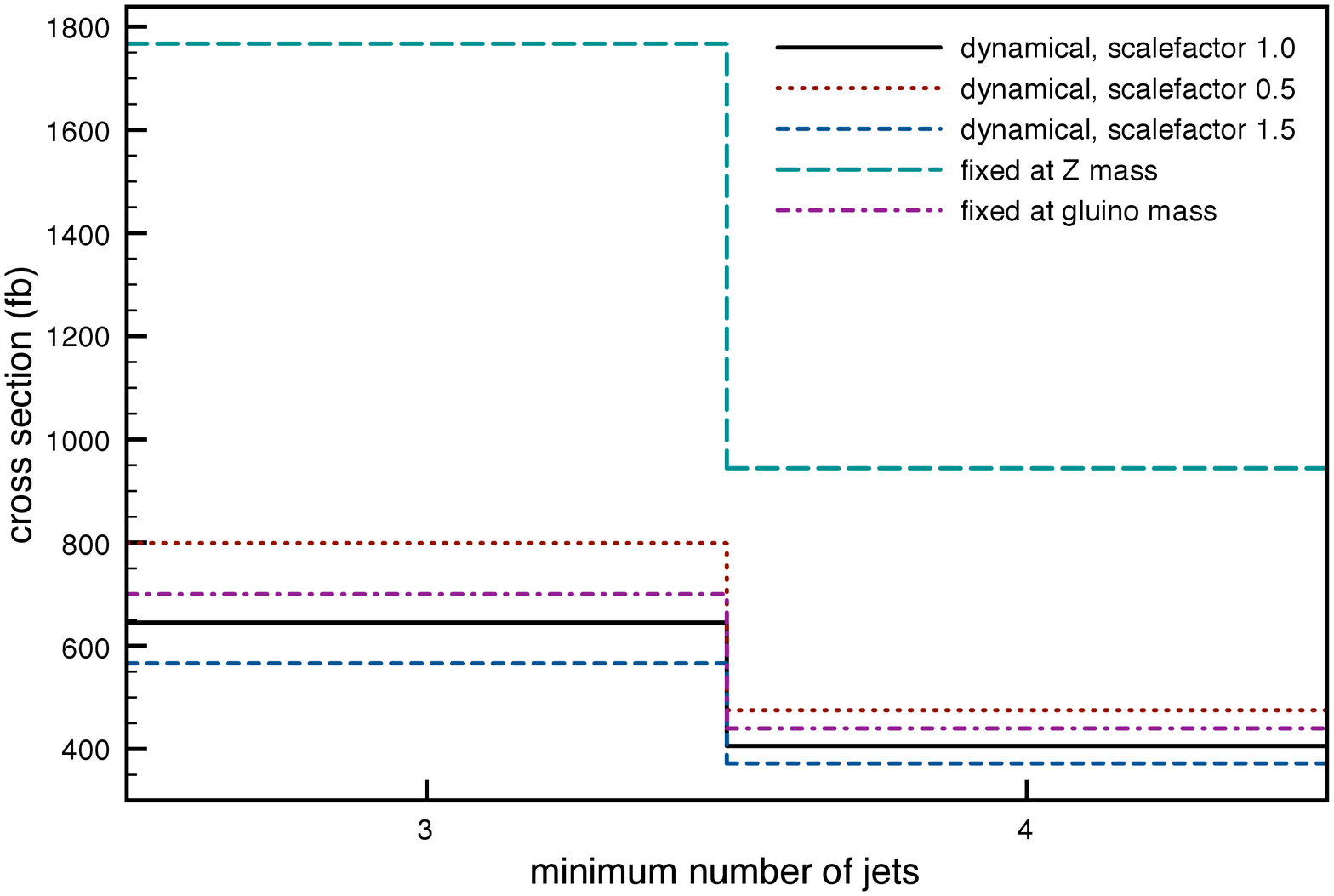}
 \caption{Tests of jet systematics on one of the benchmark models ({\bf B8}) in the dilepton channel. On the left we show the effect of a 20\% uncertainty in the jet energy determination in the early data. In the same plot, we show the effect of a sphericity cut on the signal. On the right we show the effect of different choices for the renormalization and factorization scales, where the solid black line corresponds to our analysis. The \emph{scalefactor} is an invoked rescaling factor common for all events.}
        \label{fig:syst2}
\end{figure}
%==============================================================================
The measure of significance of a signal in Eq.~(\ref{eq:detectionlimit}) assumes that the background uncertainty can be neglected. However, for the energies to be reached at the LHC a good knowledge of the standard model background cannot really be achieved before it can be \emph{measured} in the data. As stated in the introduction, our aim with this paper is not to make exact predictions for very specific models but to get an idea of what supersymmetric dark matter models can be probed using early LHC data. There are however some tests that we can perform on our simulated data to estimate how systematic uncertainties affect our results and to see how reasonable our cuts are.

One difficult problem is that of what renormalization and factorization scales to use in the event generation. Additionally, for early data the systematic uncertainty in jet energy determination can be of the order 10--20\%. 

Figure \ref{fig:syst1} shows the effects of varying the jet $p_T$ cuts by $\pm 20\%$ (which closely imitates the effect of changing the jet energy) and the choice of renormalization/factorization scale for a $t\bar{t}$ sample in the dilepton channel.\footnote{A choice of a renormalization/factorization scale scale fixed at the $Z$ mass is clearly questionable for $t\bar{t}$ and, especially, heavy sparticle production, but is anyway included here for the sake of illustration.} Figure \ref{fig:syst2} shows the effects of systematics for our benchmark model {\bf B8} correspondingly. (The systematics for all our benchmark models look similar). These systematic errors are not included in our analysis but should not alter the prediction for the detection luminosity by more than a factor of 2.

We also show that the effect on the background of cutting away events with transverse sphericity less than 0.2 drowns in the uncertainty in jet energy determination within the early data, which is why we do not include this otherwise standard requirement in our cuts. We also note that the sphericity cut can reduce the signal at least as much as the background.

We also note that while we here consistently assume a perfect knowledge of the $true$ integrated luminosity, the experimentally measured value may not agree perfectly. This could also add some contribution to the systematics in the early data.

\subsection{7 TeV reach}
%==============================================================================
\begin{figure}
        \centering
\includegraphics[scale=0.6]{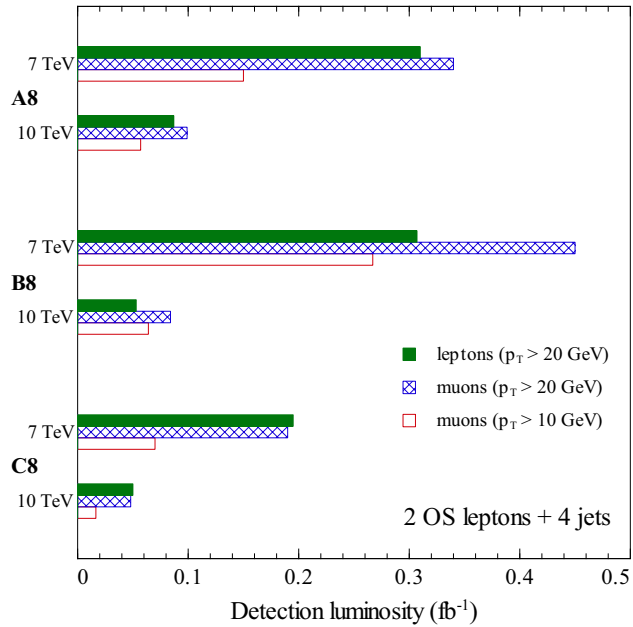}
 \caption{Integrated luminosity required for detection of the MSSM-8 benchmark models of Table \ref{table:benchmarkmodels} in the OS dilepton (electrons+muons) and OS dimuon channels, at 7 TeV as well as 10 TeV center-of-mass energy.}
        \label{fig:bmlum7TeV}
\end{figure}
%==============================================================================

We investigate the effect of lower initial beam energy by running the simulation for our MSSM-8 benchmark models at 7 TeV center-of-mass energy. To get an estimate of the background we simulate $t\bar{t}$ events at 7 TeV to a luminosity of 2.3 fb$^{-1}$ and focus on the dilepton channel for which we assume this to be the dominating background, and the OS channel i particular for which this gives a statistically representative sample.

We see from Figure \ref{fig:bmlum7TeV} that the integrated luminosity needed for detection at 7 TeV is about 3-4 times higher than at 10 TeV. This is roughly what one would naively expect from how the production cross sections scale down, together with the effect of the decreased energy available for production of initial state QCD radiation.

%%%%%%%%%%
\section{Summary and conclusions} 

We have investigated early discovery signals in the ATLAS detector at the LHC without using a cut on missing transverse energy. Compared with the earlier studies of \cite{BaernoMET,Baer2008,Baer2009}, we have instead of focusing on the constrained MSSM (or mSUGRA), studied the more phenomenological MSSM-7 and MSSM-8 models. We have also required that all our models have a neutralino that can be the bulk of the dark matter. In addition, we have used a completely different set of tools (with different systematics) than in previous studies.

\begin{figure}
       \centering
\includegraphics[scale=0.5]{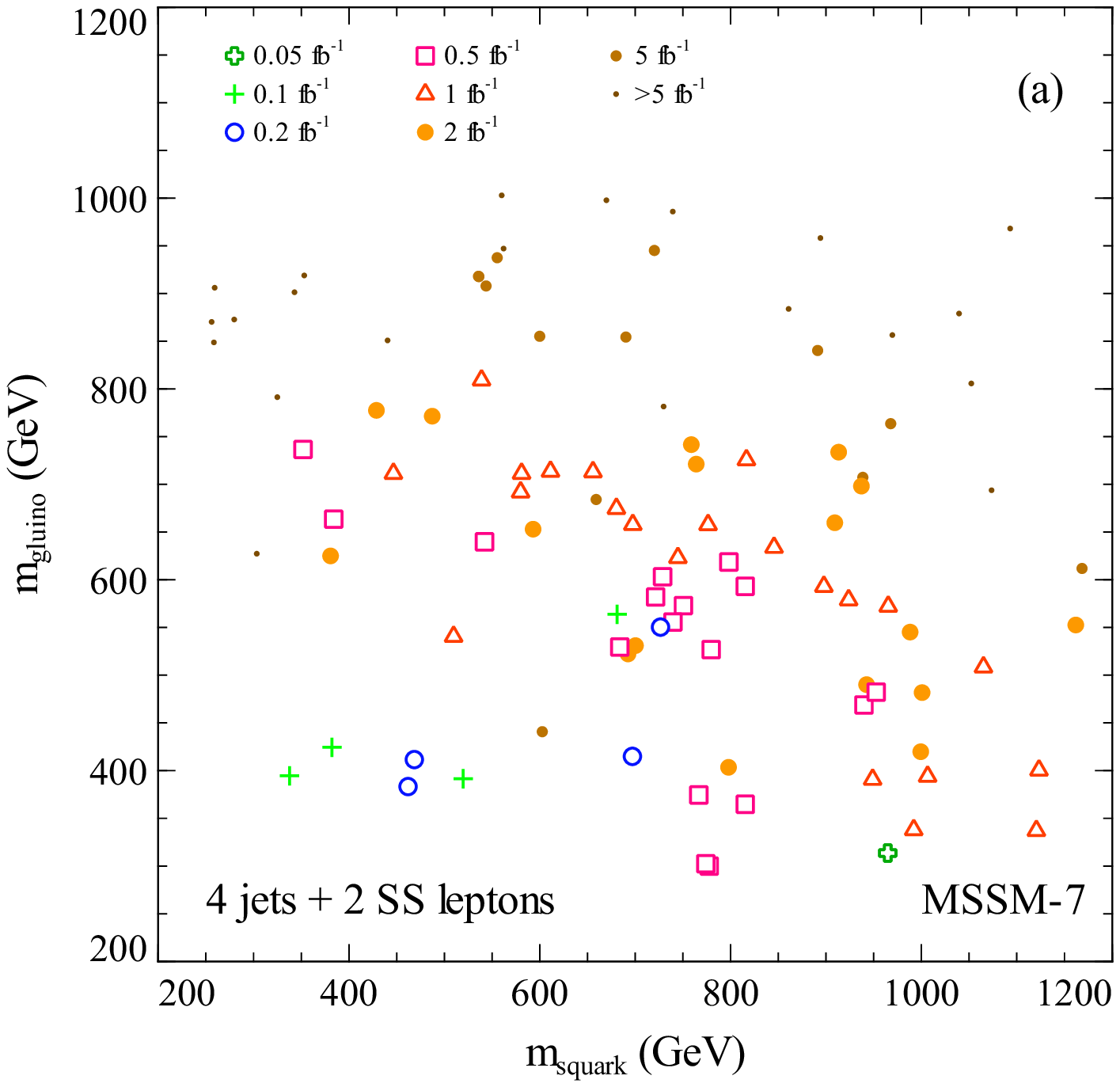}
\includegraphics[scale=0.5]{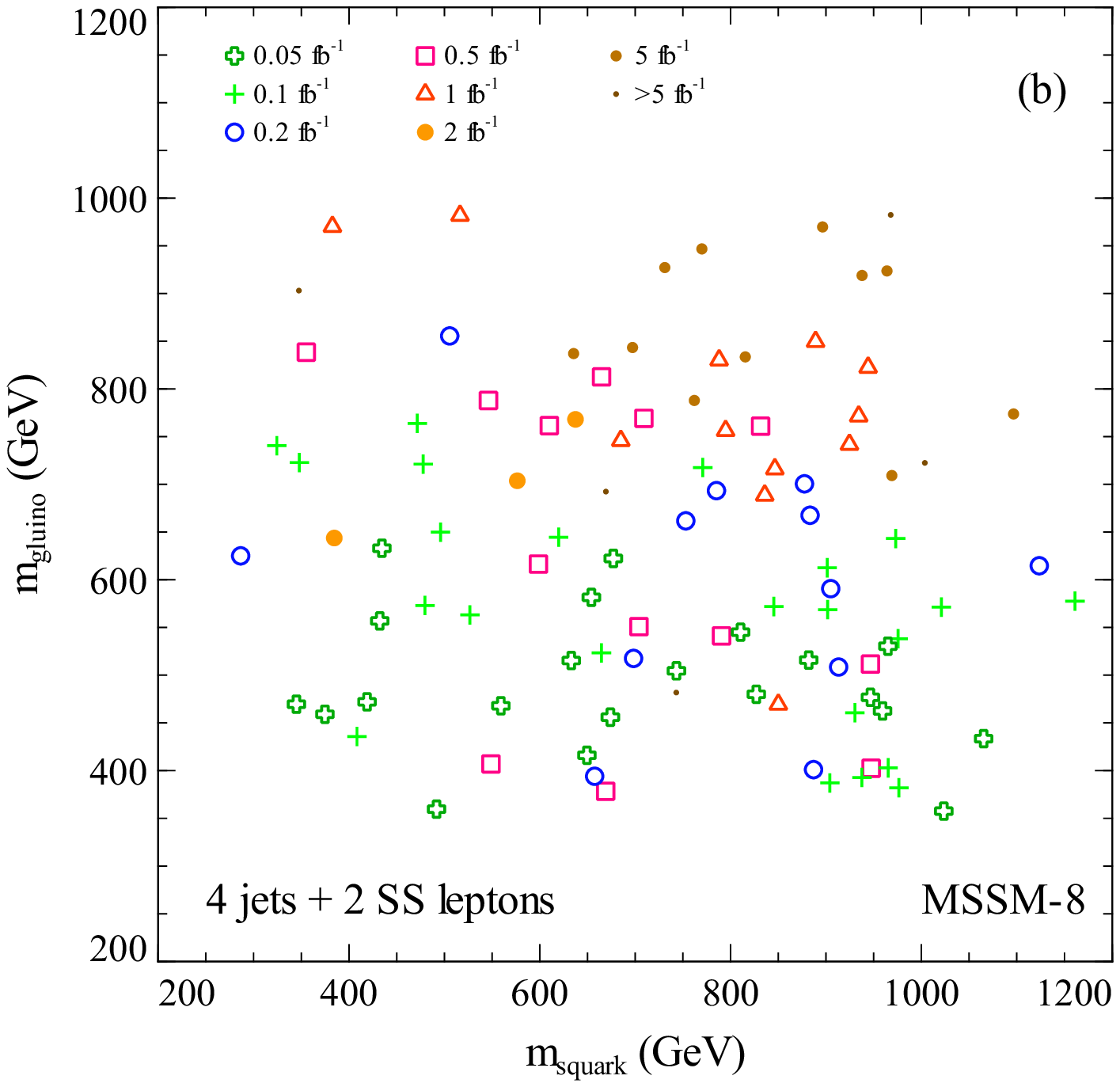}
\includegraphics[scale=0.5]{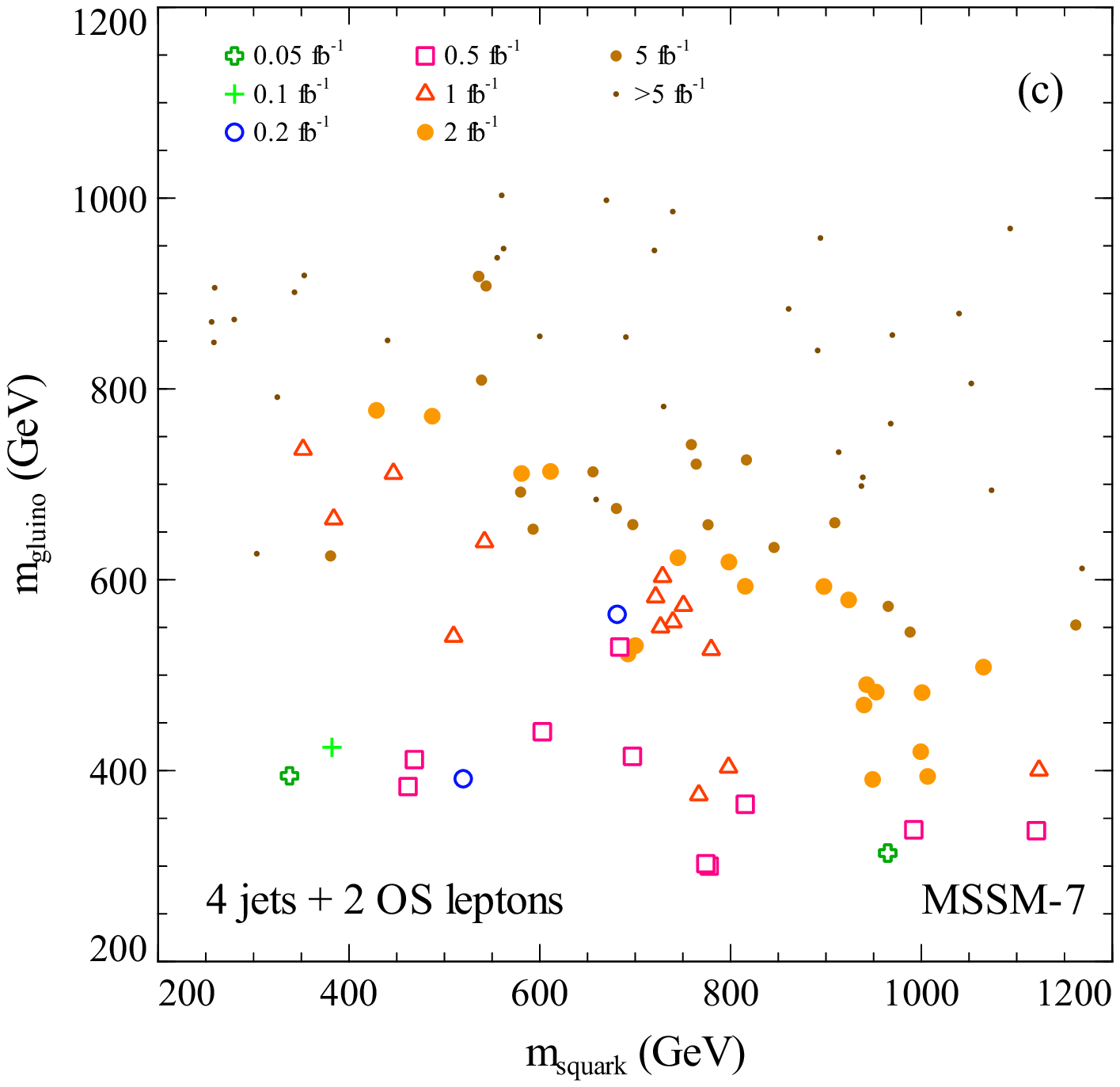}
\includegraphics[scale=0.5]{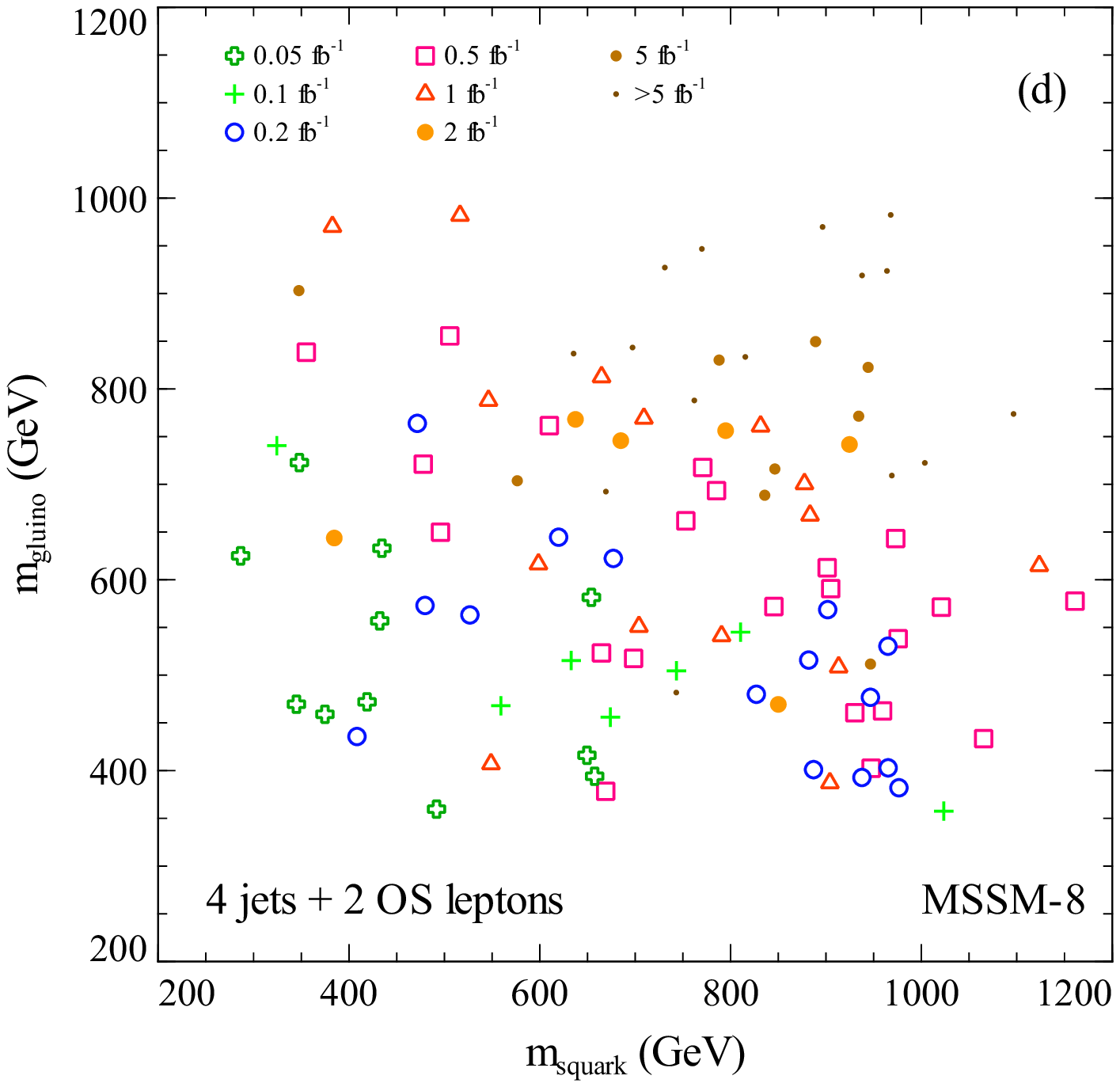}
	\caption{Integrated luminosity required for detection of (a),(c) MSSM-7 and (b),(d) MSSM-8 models in the four jets (with $p_T>100,50,50,50$ GeV) plus two leptons (each with $p_T>20$ GeV) search. The results are shown for the (a),(b) same sign and (c),(d) opposite sign dilepton channels.}
        \label{fig:4j2l}
\end{figure}

An important property to improve detectability within early LHC data, apart from lower gluino/squark masses, is the presence of light sleptons in the sparticle spectrum. In the MSSM-8 extension of the MSSM-7 model, we allow the slepton mass to be a free parameter and focus on models where sleptons are light. We conclude that a multi-jet plus dilepton search at 10 TeV, with no cut on MET, can probe supersymmetric models with gluino masses up to $\sim$ 700 GeV for arbitrary squark masses (and even slightly heavier gluinos for lighter squarks) within the first 200--300 pb$^{-1}$ period of data taking. (See first and foremost Figure \ref{fig:4j2l}.) The multi-jet plus trilepton channel is less important early on but can be used for cross-correlation at later stages.

Our results seem to be fairly consistent with those of \cite{BaernoMET,Baer2008,Baer2009} regarding background and signal estimation, although a direct comparison is often not feasible due to differences in our analyses. We find our MSSM-7/MSSM-8 models to be a bit harder/easier to detect compared to their mSUGRA models under study, something which is consistent with the notion of the mSUGRA sfermion spectrum being a mixture of those of the MSSM-7 and MSSM-8. We note that even clearer signals may potentially be achieved by extending our MSSM-8 model to allow for separate mass parameters for the individual sleptons.

It should be stressed that our imposed sets of cuts are intentionally very simple, and we expect that the LHC reach could be improved by performing more model specific searches. However, until the LHC has started collecting some data, providing a better understanding of the QCD background, large systematics remain and we find it reasonable not to go too much into detail. In fact, even though we have performed a conservative analysis in order not to overestimate the LHC reach, the systematic errors can still be quite important, and the results are to be seen as guidance.

In summary, we have confirmed that supersymmetry searches, not relying on missing energy, using the early LHC data have a substantial reach within the MSSM parameter space, and found that models with light sleptons are particularly interesting to search for early on.

\ack{We thank J. Alwall for {\tt MG/ME} support and H. Baer for discussions. We thank the Swedish Research Council (VR) for support.}

\pagebreak

\end{document}